\documentclass[twocolumn]{revtex4}

\usepackage{latexsym}

\usepackage{graphicx}
\usepackage{bm}
\usepackage{epstopdf}
\usepackage{amsmath,amsfonts,paralist}

\usepackage{subfig}

\let\a=\alpha \let\b=\beta  \let\g=\gamma  \let\d=\delta \let\e=\varepsilon
     \let\l=\lambda
                
\let\s=\sigma \let\t=\tau

  \let\r=\rho 

\def\eg{{e.g. }}

\def\EE{{\cal E}} 
\def\CC{{\cal C}}\def\FF{{\cal F}} \def\HH{{\cal H}}\def\WW{{\cal W}}
 
\def\RR{{\cal R}}  \def\OO{{\cal O}}

\def\ol{\overline}

\newcommand{\beq}{\begin{equation}}
\newcommand{\eeq}{\end{equation}}
\newcommand{\beqa}{\begin{eqnarray}}
\newcommand{\eeqa}{\end{eqnarray}}

\newcommand{\Tr}{\text{Tr}}
\newcommand{\comment}[1]{}

\date{}

\begin{document}

\title{Critical Off-Equilibrium Dynamics in Glassy Systems}

\author{Francesco Caltagirone$^{1,2}$, Giorgio Parisi$^{1,2,3}$ and Tommaso Rizzo$^{1,2}$} \affiliation{$^1$ Dip. Fisica,
Universit\`a "Sapienza", Piazzale A. Moro 2, I-00185, Rome, Italy \\
$^2$ IPCF-CNR, UOS Rome, Universit\`a "Sapienza", Piazzale A. Moro 2,
I-00185, Rome, Italy \\ $^3$ INFN, Piazzale A. Moro 2, 00185, Rome,
Italy}


\begin{abstract}
We consider off-equilibrium dynamics at the critical temperature in a class of glassy system.
The off-equilibrium correlation and response functions obey a precise scaling form in the aging regime. 
The structure of the {\it equilibrium} replicated Gibbs free energy fixes the corresponding {\it off-equilibrium} scaling functions implicitly through two functional equations.
The details of the model enter these equations only through the ratio $w_2/w_1$  of the cubic coefficients (proper vertexes) of the replicated Gibbs free energy. Therefore the off-equilibrium dynamical exponents are controlled  by the very same parameter exponent $\lambda=w_2/w_1$ that determines equilibrium dynamics.
We find approximate solutions to the equations and validate the theory by means of analytical computations and numerical simulations. 
\end{abstract} 

\date{\today} \maketitle

\section{Introduction}

The key property of glassy systems is the slowing down of the dynamics upon lowering the temperature. 
This property
makes their study so challenging both in experiments and numerical simulations.
Indeed equilibrium dynamics  becomes increasingly slow approaching the critical temperature in such a way that the relaxation time exceeds the laboratory time scale and the systems fall off-equilibrium.
This effects has its counterpart in numerical simulations where the dramatic increase of the equilibration time at low temperature strongly constrains the maximal systems size that can be equilibrated resulting in huge finite-size effects. Therefore a satisfactory theory of glassy systems must be able to characterize their off-equilibrium dynamics. On the other hand many believe that the important theoretical advances made in the context of the statics and equilibrium dynamics of these systems are useful if not essential to understand  their off-equilibrium dynamics. In particular deep connections between off-equilibrium dynamics and statics have been obtained in the study of aging \cite{Cugliandolo93,Cugliandolo94,franz99}. These studies focus on a non-trivial time-reparametrization invariance of the problem that naturally leads to a parametric ({\it i.e.} without the time) representation of two-time quantities. In this framework, the problem of the approach to equilibrium of one-time quantities, say the energy, remains open. 

It would be natural to expect that, unlike the reparametrization-invariant quantities, the corresponding dynamical exponents cannot be expressed solely in terms of quantities obtained from the statics. We will show it is possible to obtain precise results for the dynamical exponents extending some results obtained recently in the context of critical equilibrium dynamics \cite{w1w2,Calta11}.  In this paper we address the computation of the dynamical exponents for a class of glassy systems at the critical temperature in a mean field theory framework. 

The order parameter in glassy systems is typically a two-point function. In mean-field spin-glasses (SG) one considers the spin-spin correlation defined as:
\beq
C(t,s)={1 \over N} \sum_{i=1}^N \overline{\langle s_i(t) s_i(s) \rangle}
\eeq
where $N$ is the total number of spins in the system, the angle brackets mean thermal averages and the overline means average with respect to the quenched disorder \cite{Mezard87}.
At the critical temperature the equilibrium spin-spin correlation in zero external field exhibits a power-law decay in time, i.e. 
\beq
C(\tau) \propto {1 \over \t^{\nu}}
\eeq
for large values of $\tau=|t-s|$ \cite{Sompolinsky82}. 

In \cite{w1w2,Calta11} it has been argued that this behavior follows from the fact that the replicated Gibbs free energy admits the following expansion near the critical temperature $T \approx T_c$:
\beq
 G({Q}) = a_T {T-T_c \over 2} \sum_{a,b} Q_{ab}^2 
-\frac{w_1}{6} \Tr Q^3 
- \frac{w_2}{6} \sum_{a,b}Q_{ab}^3
\label{repgibb}
\eeq 
where $a_T$ is some model-dependent constant and $Q_{ab}$ is a replicated version of the two-point order parameter. 
Furthermore it has been shown that the so-called parameter exponent $\lambda$  which determines $\nu$ through the following relationship:
\beq
{\Gamma^2(1-\nu) \over \Gamma(1-2 \nu)}=\lambda \ ,
\label{lanu}
\eeq
is equal to the ratio between the effective coupling constants $w_2$ and $w_1$:
\beq
\lambda={w_2 \over w_1}\ .
\label{law}
\eeq 

The fact that equilibrium dynamics follows from the static replicated Gibbs free energy makes it rather universal. Indeed, in the Landau sense, one can argue that the structure of the Gibbs free energy near the transition depends solely on the symmetries of the problem and therefore could be the same for quite different models.
Notable examples of models  whose replicated Gibbs Free energy admits the expansion (\ref{repgibb}) near the critical temperature are the Sherrington-Kirkpatrick (SK) model in zero field, various spherical $p$-spin fully-connected models in zero field, the Potts SG with $p=3$ (both fully connected and on the Bethe lattice) and instances of the so-called $M-p$ models \cite{Calta10} for appropriate values of the parameters $M$ and $p$. 
We also expect that the structure of the mean-field free energy remains the same when the above models are defined on finite dimensional lattices above the upper critical dimension $D=6$. The corresponding transition has been also encountered in the study of schematic Mode-Coupling-Theory (MCT) models for supercooled liquids. In the original MCT literature it was called a type A transition while in the modern terminology is called a degenerate $A_2$ singularity (see \cite{Gotze}, pag. 228).  We recall that in the context of MCT the exponent $\nu$ is usually called $a$. In the SG literature the corresponding transition is called a continuous transition in zero external field. It has to be contrasted with the continuous transition in a field and with the discontinuous transition whose replicated Gibbs free energy contain additional terms with respect to (\ref{repgibb}) \cite{w1w2}.

In this paper we will consider the correlation and response functions $C(t,s)$ and $R(t,s)$ defined as
\beq
R(t,s) \equiv \sum_i \overline{\delta \langle s_i(t) \rangle/\delta h_i(s)}
\eeq
where the $h_i(s)$'s are small auxiliary time-dependent external fields that enters in the Hamiltonian as $\sum_{i}s_i h_i(s)$ and are set to zero after taking the derivative.
We will discuss the behavior of $C(t,s)$ and $R(t,s)$ at the critical point upon dynamical evolution starting from random initial configurations at time $t=0$. 
This is equivalent to an instantaneous quench from $T=\infty$ to the critical temperature $T=T_c$. We will focus on the so-called aging regime in which both $t$ and $s$ are large.

We will first describe the spherical $p$-spin model with $p=2$ which admits a full analytical solution. Interestingly enough, a dynamical computation for the SK model (reported in Appendix \ref{app:sk}) shows that the off-equilibrium dynamical exponents of these two models are the same. 
Guided by these findings, we will argue that {\it for all models that have a replicated Gibbs free energy with the above structure (Eq. (\ref{repgibb})) the correlation and response functions in the aging regime are described by appropriate scaling functions from which various dynamical exponents can be extracted. Remarkably the scaling functions depend on the details of the model only through the very same parameter exponent $\lambda=w_2/w_1$ that controls critical equilibrium dynamics}.  

Technically speaking, the above replicated action describes continuous SG transitions characterized by the simultaneous vanishing of the replicon, longitudinal and anomalous eigenvalues. The case in which only the replicon eigenvalue vanishes requires some non-trivial modifications and is left for future work. More physically, we note that action (\ref{repgibb}) is a special case of a more general action that should contain at the quadratic level also terms of the form $m_2 \sum_{abc} Q_{ac}Q_{ab}$ and $m_3 \sum_{abcd} Q_{ab} Q_{cd}$ \cite{w1w2}. The coefficients of these terms vanish if the Hamiltonian of the model display additional symmetries (besides replica-symmetry), for instance time-reversal for Ising spins or the Potts symmetry for spins with $p$-states. Therefore an important case which cannot be described by the present theory is the SK model in a field. 

The theory yields equations from which in principle the scaling functions can be computed for any value of the parameter exponent $\lambda$. At present we have no analytical solution of the equations for general values of $\lambda$ but we have devised an approximation scheme that yields consistent estimates of the scaling functions and exponents for not too large values of $\lambda$. 

Novel predictions for the approach to equilibrium of one-time quantities can also be obtained, notably the energy and magnetization decay that are typical quantities measured in numerical simulations. It is found that the energy approaches its equilibrium value at infinite time according to:
\beq
E(t)-E_{\infty} \propto {1 \over t^{2 \nu}}
\eeq 
meaning that the dynamical exponent of the off-equilibrium decay of the energy is two times the exponent of the equilibrium correlation $\nu$.
The decay of the remanent magnetization or equivalently the decay to zero of the correlation between the initial random configuration and the configuration at a large time $t$ is given by:
\beq
m_R(t) \propto C(0,t) \propto {1 \over t^\delta} \ .
\eeq
According to the theory, the exponent $\delta$ obeys the following relationship
\beq
\delta = \alpha + \nu
\eeq
where $\alpha$ is a novel exponent associated to the behaviour at small argument of the scaling functions for the correlation and response. Specializing to the Sherrington-Kirkpatrick model in zero external field the theory yields $\nu=1/2$ and $\alpha=3/4$ leading to a $1/t$ decay of the energy and to an exponent $\delta=5/4$ consistently with a previous direct analisys and numerical simulations \cite{Ranieri97}.

The predictions of the theory have been validated in two ways. We considered the $2+3$ class of spherical Spin-Glass models where the parameter $\lambda$ can be tuned between $0$ and $1$ and solved the exact off-equilibrium dynamical equations by means of a power series expansion at small times. The method allows to control precisely the region of moderately small values of $\lambda$ where the decay exponents are not too small. In this region we have found a very good agreement with  the results coming from the numerical solution of the universal equations \footnote{One could have also studied the same equations using adaptive algorithms as in \cite{Kim01,Andre06,Miyaz07} }. 
We have also performed a numerical simulation on the fully connected three-states Potts Spin-Glass at the critical temperature. In this case $\lambda=1/2$ and we have again found a very satisfying agreement with the predictions of the theory for the decay of the energy and for the  various dynamical exponents obtained from the numerical solutions of the universal equations. 

 The paper is organized as follows.
In section \ref{sec:scenario} we will present the general scenario for the off-equilibrium critical behavior of the class of systems considered. 
In section \ref{sec:spherical} we will give a detailed treatment of the off-equilibrium dynamics of the spherical $2$-spin model  showing explicitly that it follows the general scenario.
In section \ref{sec:general} we will study off-equilibrium dynamics in the quasi-static limit and argue that the above scenario applies to all systems whose replicated Gibbs free energy admits the expansion (\ref{repgibb}) following the  procedure of \cite{w1w2}.
In Section \ref{sec:numsol} we will present a method to solve numerically the universal equations describing the correlation and response function and give the result of this analysis.
In Section \ref{sec:valther} we will validate the theory presenting results from an off-equilibrium numerical simulation on the three-states Potts SG and on the solutions of the exact dynamical equations for the spherical models.
In Section \ref{sec:conclusion} we will give our conclusions.
Various computations and results will be presented in the appendices.

\section{The General Scenario}
\label{sec:scenario}

We  consider a general scenario in which off-equilibrium critical dynamics can be characterized by the following three regimes depending on the value of $t$ and of $\tau \equiv t-s$:
\begin{itemize}
\item The equilibrium regime corresponding to $t \gg 1$  while $|t-s| \ll t$. In this regime the two functions become equal to their equilibrium limit:
\beq
\begin{split}
&C(t,s) \approx C_{eq}(t-s)\\ 
&R(t,s) \approx R_{eq}(t-s) \\
&\text{for }\  \  t \gg 1\, ,\,\,|t-s| \ll t 
\end{split}
\eeq 
the precise form of the function $C_{eq}(\tau)$ at small time differences $\tau$ depends on the microscopic details of the model and of the type of dynamics. However as we said before the exponent $\nu$ of the long time power-law decay at criticality depends only on the parameter $\lambda=w_2/w_1$ through Eq. (\ref{lanu}).

\item The aging regime in which $t \gg 1$ and $\tau$ is also large such that $s/t=a$ remains finite while $t$ tends to infinity. 
Recall that in the present paper we are considering the aging regime {\it at} the critical temperature while for a study 
of the aging dynamics {\it below} the critical temperature we refer the reader to \cite{Cugliandolo93,Cugliandolo94,franzparisi2012}. 

 In this regime we have:
\beqa
C(t,a \, t) &  = & {c \over t^{\nu}} \CC(a)+o(t^{-\nu})
\\
R(t,a \, t) & = &  {c \over t^{\nu+1}} \RR(a)+o(t^{-\nu-1})
\eeqa
where the exponent $\nu$ is the same of the equilibrium regime and $c$ is a model-dependent constant prefactor.  

The two scaling functions $\CC(a)$ and $\RR(a)$ are determined by two quadratic equations that depend solely on $\lambda$.
In order to write the equations it is convenient to define:

\beqa
\CC_{eq}(a) & \equiv & {1 \over \nu (1-a)^{\nu}}
\\
\RR_{eq}(a) &  \equiv & {1 \over  (1-a)^{\nu+1}}
\eeqa
The behaviour of the two functions $\CC(a)$ and $\RR(a)$ in the limit $a \rightarrow 1$ matches the equilibrium behaviour and we have:
\beqa
\CC(a) & \xrightarrow[a \rightarrow 1]{} & \CC_{eq}(a)
\\
\RR(a) & \xrightarrow[a \rightarrow 1]{} & \RR_{eq}(a)
\eeqa
The equations for $\CC(a)$ and $\RR(a)$ read:

\begin{widetext}

\beq
\nonumber
\begin{split}
&\int_0^a \, a^{-\nu} \RR(b) \CC\left( \frac{b}{a} \right) \, db + \int_0^a \left[  \CC(b) a^{-\nu -1} \RR \left(\frac{b}{a}\right) - \CC(a) a^{-\nu -1} \RR_{eq} \left( \frac{b}{a} \right) \right] \, db - \\
& -  \CC(a) a^{-\nu} \CC_{eq}(0) - \CC(a) \CC_{eq}(a) + \int_a^1 \left[  \RR(b) b^{-\nu} \CC \left(\frac{a}{b}\right) - \RR_{eq}(b) \CC(a)\right] \, db + \l \CC^2(a)=0
\end{split}
\eeq
and
\beq
\nonumber
\begin{split}
& \int_a^1 \, \left\lbrace \RR(b) b^{-\nu -1} \RR \left( \frac{a}{b} \right) - \RR(a) \left[ \RR_{eq}(b) +  \RR_{eq} \left( \frac{a}{b} \right)\right] \right\rbrace \, db - \\
& - 2 \RR(a)\CC_{eq}(a) + 2\l \RR(a) \CC(a)=0
\end{split}
\eeq

\end{widetext}

In the case $\lambda=0$ the solution of the above equations is: 
\beqa
\CC(a) & = &  \frac{ 4 a^{3/4}}{ \,(1+a) (1-a)^{1/2} }
\label{solf1}
\\
\RR(a) & = & \frac{a^{-1/4}}{ \, (1-a)^{3/2} }
\label{solf2}
\eeqa
The above solution correspond to the spherical model with $p=2$, in this case $C(t,a t)$ and $R(t,at)$ can be computed explicitly and the constant $c$ turns out to be equal to $1/(2 \pi^{1/2})$. 
For general values of $\lambda$ we cannot exhibit explicitly the solution of the above equations.
However we expect that the two solutions at small values of $a$ have a power law behaviour controlled by a single exponent $\alpha$ according to:
\beqa
\CC(a) & \propto & a^{\alpha}, \,\,\,\,\,\, a \approx 0
\\
\RR(a) & \propto & a^{\alpha-1}, \,\,\,\,\,\, a \approx 0
\eeqa

\item The regime in which $t \gg 1$ while $s$ is finite. In this case we have:
\beqa
\label{finitetw}
C(t,s) \simeq {1 \over t^{\delta}} c(s) 
\\
R(t,s) \simeq {1 \over t^{\delta}} r(s)
\eeqa
Similarly to the equilibrium case, the precise form of the two functions $c(s)$ and $r(s)$ at finite $s$ depends on the details of the model.
However, by means of matching arguments, their large-$s$ behavior and the value of the exponent $\delta$ can be inferred from the small $a$ behaviour of the functions $\CC(a)$ and $\RR(a)$ of the aging regime and therefore are fixed by the parameter $\lambda$.
More precisely we expect that
\beqa
c(s) & \propto & s^{\alpha}, \,\,\, s \gg 1
\\
r(s) & \propto & s^{\alpha-1}, \,\,\, s \gg 1
\eeqa 
and 
\beq
\delta=\alpha+\nu
\eeq

\end{itemize}
The exponent $\d$ is the same of the long time power-law decay of the {\it remanent magnetization} 
\beq
m_R^{(t_w)}(t)=\int_0^{t_w} R(t,s) \, ds
\eeq
for finite waiting times $t_w$. In fact it is straightforward that for $t\gg 1$ and $t_w/t \ll 1$
\beq
m_R(t) \propto R(t,0)
\eeq
Since our interest will be in the asymptotics and, in particular, in the exponent $\d$, we will make no distinction 
between the two quantities and we will simply refer to $R(t,0)$ as the ``remanent magnetization'' throughout the paper. \\
\indent We will also obtain a general prediction on the off-equilibrium behavior of the energy.
The form of the Replicated Gibbs free energy (\ref{repgibb}) tells us that deviations of the energy from its equilibrium value are controlled in replica space by the quantity $\sum_{a,b} Q_{ab}^2 $, from this one can argue that in off-equilibrium dynamics the energy approaches its equilibrium value in the following way:
\beq
E(t)=E_{\infty}+{a_E(\nu) \over t^{2 \nu}}
\eeq
meaning that the energy has a power-law relaxation to equilibrium with an exponent two times $\nu$.
The coefficient $a_E$ can be expressed in terms of the model dependent constants $a_T$ and $c$ and by means of the functions $\CC(a)$ and $\RR(a)$ as:
\beq
\begin{split}
&a_E(\nu)=a_T \, c^2 \left[ \int_0^1 \left[ \CC(a)\RR(a)-\CC_{eq}(a)\RR_{eq}(a) \right] -{1 \over \nu^2}  \right] 
\end{split}
\label{ener}
\eeq

The most interesting features of the present scenario is that many of the dynamical off-equilibrium critical exponents are determined by the very same exponent parameter $\lambda$ controlling the equilibrium dynamics. In particular the exponents $\alpha$ and $\delta$ (through $\delta=\alpha+\nu$) are both determined by the universal aging-regime equations for $\CC(a)$ and $\RR(a)$. This type of equations is not well studied in the literature and it is not clear to us if it is possible to find an explicit analytical solution  when $\lambda\neq 0$. Due to the singular nature of the solutions it is also not simple to solve them numerically, nevertheless in Section \ref{sec:numsol} we will present a variational scheme that appears to give consistent results. The method uses appropriate trial functions for $\CC(a)$ and $\RR(a)$ which are fixed minimizing the square of the deviations of the exact equations on a set of points between zero and one. The procedure requires that the integral equations are recast in order to render the singularities in the numerical integrals harmless. Once this is achieved integrating by parts, a standard Gauss-Newton minimization scheme appears to converge rather fast. In this respect we believe that the problem at the numerical level is essentially solved: having more precise results than those we will present is only a matter of computational time and numerical precision.

\section{Spherical $2$-spin model}
\label{sec:spherical}
The off-equilibrium dynamics of the fully-connected spherical $2$-spin model \cite{Kosterlitz76,Ciuchi88}, has been 
solved exactly below the critical temperature in \cite{cugliandolo_dean} through a projection on the eigenvalues 
of the (random) interaction matrix.

In this Section we  study the off-equilibrium dynamics {\it at} the critical
temperature starting from a random
configuration at time zero and, in particular, the asymptotic long time behaviour. We will basically follow the approach and the notation of \cite{dedom_giardina}. Here we give the main results, while the details of the computation can be found in Appendix \ref{app:spherical}.\\
The Hamiltonian of the model is given by
\beq
\HH=-\frac{1}{2}\sum_{ i\neq j } J_{ij} s_i s_j
\eeq
where the spins are continuous variables satisfying a global spherical constraint
\beq
\sum_{i=1}^N s_i^2 = N
\eeq
and the couplings are independent random variables following a Gaussian distribution with zero mean 
\beq
P(J_{ij})=\frac{1}{\sqrt{2\pi J^2}} \exp \left( -N\frac{J_{ij}^2}{2J^2} \right)
\eeq
It can be shown \cite{dedom_giardina} that the eigenvalue density distribution of the random interaction matrix follows the well 
know Wigner semi-circle law in the thermodynamic limit and is sample independent at leading order, namely 
\beq
\r(\mu)=\frac{1}{2\pi J^2} \sqrt{4J^2-\mu^2} \,\,\,\,\,\,\,\,\,\, |\mu | \leq 2J
\eeq
The critical temperature of the model is given by \cite{dedom_giardina}
\beq
\frac{1}{T_c}= \int_{-2J}^{2J} \, d\mu \, \r(\mu) \, \frac{1}{2J-\mu}
\eeq
As already pointed out, the projected Langevin equation corresponding to this system can be solved exactly and 
the correlation and the response functions can be expressed in terms of a function $\gamma(t)$ in the following way
\beq
\label{corre}
\begin{split}
C(t,s)=\int_{-2J}^{2J} \, d\mu \, \rho(\mu) \left[ s_{\mu}^2(0) e^{-(2J-\mu)(t+s) } \g (t) \g (s) \right.\\
 +  \left. 2T \int_0^{\text{min} (t,s)} dt' \, e^{-(2J-\mu)(t+s-2t')} \frac{\g (t) \g (s)}{\g^2(t')}\right]
\end{split}
\eeq
\beq
\label{resp}
R(t,s)= \int_{-2J}^{2J} \, d\mu \, \rho(\mu) \, e^{-(2J-\mu)(t-s) } \frac{\g (t)}{\g (s)}
\eeq
where $\g (t)$ satisfies the integral equation (\ref{constraint}) given in the Appendix, considering the definition (\ref{bigD}). \\ 
Through an asymptotic analysis of the integral equation (see Appendix \ref{app:spherical}) it can be seen that, {\it at} the critical temperature 
$T=T_c=1$, the laeding behaviour of $\g$ for large times is
\beq
\label{gamma}
\g (t)\simeq 2^{3/4} \pi^{1/4} t^{1/4}
\eeq

It has been shown in \cite{zippold} that in the low temperature phase ($T<T_c $) and for large waiting times $s$ (or $t_w$), three different time-scales can be identified: two time-scales are more evident and were already discussed in \cite{cugliandolo_dean}  while the third one is more subtle. 

The first regime is the equilibrium one where $s \rightarrow \infty$, $\t /s \equiv (t-s)/s << 1$, in which FDT holds. The second regime is the 
{\it aging} one, where $s \rightarrow \infty$ and $\t \propto s$ and the scaling variable becomes the ratio $t/s$.

The third time-scale is intermediate and corresponds (below $T_c$) to the plateau preceding the {\it aging} part. This time-scale 
is a function of the waiting time and, more precisely, it corresponds to $\t \sim s^{4/5} << s$ (see \cite{zippold} for the details).

As we will see, this third time-scale is absent at the critical temperature, basically because there is no plateau in the correlation. 

We consider now the system at criticality ($T=T_c$) and we introduce two scaling functions in the regime where $t, s\rightarrow \infty$ with $s/t=a$ ({\it aging regime}). From Eq.s (\ref{corre}) and (\ref{resp}), considering 
the asymptotic behavior of $\gamma (t)$ we obtain 
\beq
\label{corr_aging1}
C(t,at) \simeq \frac{2 a^{3/4}}{\pi^{1/2} (1+a) (1-a)^{1/2} t^{1/2}}
\eeq
\beq
\label{resp_aging1}
R(t,at)\simeq \frac{a^{-1/4}}{2 \pi^{1/2} (1-a)^{3/2} t^{3/2}}
\eeq

From Eq. (\ref{corr_aging1}) it is clear that this regime describes the correlation near its equilibrium value which is 0, in fact there is 
a prefactor $t^{-1/2}$ ensuring that $C$ is small given any $t$ large. This is due to the fact that {\it at} the critical temperature there 
is no plateau and no aging, at difference with the case below $T_c$ where the correlation function 
stays close to the plateau in a regime $\t \sim t_p(s)$ which is intermediate between equilibrium and aging. 

In the large time equilibrium regime we consider $C(s+\t, s)$ with $s\rightarrow \infty$, $\t >> 1$ and $\t/s \rightarrow 0$. This means that we have to take 
first the $s \rightarrow \infty$ limit and then take $\t$ very large. Discarding the corrections of the prefactor in $\t/s$, the leading order gives 
\beq
\label{corr_eq}
C(\t) \simeq \frac{1}{\pi^{1/2} \t^{1/2}}
\eeq
\beq
\label{resp_eq}
R(\t) \simeq \frac{1}{2 \pi^{1/2} \t^{3/2}}=\frac{dC}{dt} \, ,
\eeq
which is consistent with the fact that in this regime FDT must hold.

Finally we consider a different situation, namely 
$t\rightarrow \infty$ and $s \sim 1$, and using again the long-time behavior of the function $\g(t)$ we easily find
\beq
\label{corr_smalltw}
C(t,s) \simeq c(s) t^{-5/4}
\eeq 
\beq
\label{resp_smalltw}
R(t,s) \simeq r(s) t^{-5/4} 
\eeq 
with 
\beq
\begin{split}
&c(s)\equiv \g (s) \left[ \frac{1}{(2\pi)^{1/4}}+\left( \frac{8}{\pi} \right)^{1/4} \int_0^{s} \frac{1}{\g^2(t')} dt' \right] \\
&r(s)\equiv \frac{1}{(2\pi)^{1/4} \g (s)}
\end{split}
\eeq
Note that at finite $s$, the correlation and the response exhibit the same power law behavior for large $t$ with different non-universal prefactors, $c(s)$ and $r(s)$ 
respectively, depending on $s$ in a non-trivial way.
Fixing $s=0$ we have instead that the two prefactors become exactly the same, as it should be, since the two functions $C(t,0)$ and $R(t,0)$ are 
indeed identical, as can be seen from equations (\ref{corre}) and (\ref{resp}):
\beq
C(t,0)=R(t,0)=\int_{-2J}^{2J}\, d\mu \,\rho(\mu) e^{-(2J-\mu)t} \gamma(t) 
\eeq

So far we have computed separately the asymptotic behavior of the correlation and response in three different regimes, starting from 
their closed analytic form. On the other hand, supposing that we knew only the scaling in the {\it aging} regime given in Eq.s (\ref{corr_aging1}) and (\ref{resp_aging1}), 
the scaling in the other regimes could have been derived through matching arguments. \\ 
The long waiting time behavior of (\ref{corr_smalltw}) and (\ref{resp_smalltw}) must match the behavior of (\ref{corr_aging1}) and (\ref{resp_aging1}) close to $a=0$, in fact
\beq
\begin{split}
&c(s)\simeq\frac{2}{\pi^{1/2}}s^{3/4}\\
&r(s)\simeq\frac{1}{2\pi^{1/2}}s^{-1/4}
\end{split}
\eeq
which could have been obtained from (\ref{corr_aging1}) and (\ref{resp_aging1}) taking the leading order for small $a$ and then substituting $a=s/t$. 

Moreover the asymptotic behavior of (\ref{corr_eq}) and (\ref{resp_eq}) must match (\ref{corr_aging1}) and (\ref{resp_aging1}) close to $a=1$. Again, one can derive it taking the leading order in $a \approx 1$ and substituting $t(1-a)=\t$. \\
We found convenient to use this parameter $a \in [0,1]$ but the same results can be obtained considering the more common $b=t/t_w$ with $b \in [1,\infty)$. 
In this case the scaling functions read
\beq
\label{corr_aging2}
C\left(b t_w,t_w \right) \simeq \frac{2 b^{1/4}}{\pi^{1/2} (b+1) (b-1)^{1/2} t_w^{1/2}}
\eeq
\beq
\label{resp_aging2}
R\left(b t_w,t_w \right) \simeq \frac{b^{1/4}}{2 \pi^{1/2} (b-1)^{3/2} t_w^{3/2}}
\eeq
and the matching with the finite-waiting-time regime is acheived for $b \rightarrow \infty$. \\
\indent The case of the spherical $2$-spin model is particularly simple and the dynamics can be solved analytically in all details, while this is not 
true in general for models displaying a continuous transition. In the next section we generalize these results using an effective field-theory approach. In particular, we show that, for a generic continuous model, the exponents of the relaxation of one-time quantities (\eg energy and remanent magnetization) are ruled by the exponent parameter $\l$.

\section{General Systems}
\label{sec:general}
In this section we argue that the scenario for the off-equlibrium dynamics described in Section \ref{sec:scenario} holds for any model whose replicated Gibbs free energy admits an expansion of the form (\ref{repgibb}) near the critical temperature.
We will basically apply the same arguments used in Refs. \cite{w1w2,Calta11} in an equilibrium context.

We consider a super-field formulation of dynamics in which one obtains a dynamical equation of state for the correlation and response.
In the so-called Fast Motion (FM) limit, microscopic dynamics is infinitely fast and the system reaches equilibrium instantaneously.
In this limit the correlation and response are given by the equilibrium solution:
\beq
Q_{eq,FM}(1,2)=C_{eq}(0)\delta(1,2)
\eeq
where $1$,$2$ are superfield variables.
Following \cite{w1w2} we argue that in the large time limit off-equilibrium dynamics can be described expanding the dynamical equation of state around the FM solution. This corresponds to the assumption that,  on large time scales, we are essentially in a quasi-equilibrium situation in which all one-time quantities are near their equilibrium value.
The same arguments of \cite{w1w2} lead to the conclusion that, in this limit, the dynamical equation of state reduces to the following equation for $\delta Q(1,2) \equiv Q(1,2)-Q_{eq,FM}(1,2)$:
\beq
w_1 \int d2 \delta Q(1,2) \delta Q(2,3) + w_2 \delta Q(1,3)^2=0\ .
\eeq
where the coefficients $w_1$ and $w_2$ are the same of the static replicated Gibbs free energy (\ref{repgibb}).
Note that there are no explicit time derivatives in the above equation as well as in the equilibrium case.
We have also set to zero the first order terms assuming that we are at the critical temperature.  

Following \cite{w1w2}, we can rewrite the above equation explicitly in terms of the response and correlation
function, we obtain the following two equations \footnote{Here and in the following we define the ``response'' $R(t_1,t_2)$ as $T$ times the actual response so that the temperature does not appear explicitly in the equations}:
\begin{widetext}
\beq
\int_0^{t_1}R(t_1,t_2) C(t_2,t_3)dt_2+\int_0^{t_3}dt_2 R(t_3,t_2)C(t_2,t_1)-2
C_{eq}(0)C(t_1,t_3)+{w_2 \over w_1} C(t_1,t_3)^2=0
\eeq
\beq
\int_{t_3}^{t_1}R(t_1,t_2) R(t_2,t_3)dt_2-2
C_{eq}(0)R(t_1,t_3)+2 {w_2 \over w_1} C(t_1,t_3) R(t_1,t_3)  =0
\eeq
\end{widetext}
Similarly to what we did in the equilibrium treatment we want to get
rid of the model dependent constant $C_{eq}(0)$, this can be done using Fluctuation-Dissipation Theorem {\it e.g.} $C_{eq}(0)=
\int_{-\infty}^{t_1^+} R_{eq}(t_1,t_2)dt_2 $.
By means of some manipulations we can rewrite the equations as:
\begin{widetext}
\beq
\begin{split}
&\int_0^{t_3} R(t_1,t_2)C(t_3,t_2) dt_2 + \int_0^{t_3}  [C(t_1,t_2) - C(t_1,t_3)] R(t_3,t_2) dt_2-  C(t_1,t_3) C_{eq}(t_3,0)+ \\
&+ C(t_1,t_3)\left[  \int_0^{t_3} \left[  R(t_3,t_2) - R_{eq}(t_3,t_2)\right] dt_2 +C(t_1,t_3)  - C_{eq}(t_1,t_3) +  \int_{t_3}^{t_1} [R(t_1,t_2) - R_{eq}(t_1,t_2)]  dt_2\right]\\
& + \int_{t_3}^{t_1} R(t_1,t_2) [C(t_2 ,t_3) - C(t_1,t_3)]  dt_2 + \left( \frac{\omega_2}{\omega_1} - 1  \right) C^2(t_1,t_3)=0
\label{CORRE}
\end{split}
\eeq
\beq
\begin{split}
&\int_{t_3}^{t_1} [R(t_1,t_2) - R(t_1,t_3)][R(t_2,t_3) - R(t_1,t_3)] dt_2 \\
+ & R(t_1,t_3) \int_{t_3}^{t_1}  [R(t_1,t_2) - R_{eq}(t_1,t_2)]  dt_2 + R(t_1,t_3) \int_{t_3}^{t_1}  [R(t_2,t_3) - R_{eq}(t_2,t_3)] dt_2\\
 - & 2 R(t_1,t_3) C_{eq}(t_1,t_3) - (t_1-t_3) R(t_1,t_3)^2 + 2\frac{\omega_2}{\omega_1} C(t_1,t_3) R(t_1,t_3)=0
\end{split}
\label{RESPO}
\eeq
\end{widetext}
The above equations describe the correlation and response in the region where both $C(t_1,t_2)$ and $R(t_1,t_2)$ are small. This means, in particular, that times must be large but also well separated. As we will see they were written in a form that allows to take the
large time limit inside the integrals keeping the result finite.
In order to proceed, we note that for $t_3 \rightarrow t_1$ we expect that $C(t_1,t_3)$ and $R(t_1,t_3)$ tend to their equilibrium value. In this limit we expect that the above equations reduce to the critical equilibrium dynamical equations considered in \cite{w1w2}.
The above equations were rearranged in such a way that the critical equilibrium equations correspond to last line of (\ref{CORRE}) that at criticality  admits the solution
\beq
\begin{split}
&C_{eq}(t_1,t_3)={1 \over \nu |t_1-t_3|^{\nu}} \\  
&R_{eq}(t_1,t_3)={1 \over  |t_1-t_3|^{\nu+1}}
\end{split}
\label{criteq}
\eeq   
Thus the fact that we are off-equilibrium is encoded by the presence of the terms in the first two lines of Eq. (\ref{CORRE}).
If we plug the critical equilibrium solution  (\ref{criteq}) in (\ref{CORRE}) we find that the second line gives trivially a vanishing contribution while the first line yields a term $1/(t_1 t_3)^{\nu}$. This term can be treated as a small correction to the last line which is of order $1/|t_1-t_3|^{2\nu}$ as long as $|t_1-t_3|\ll t_1$ and this corresponds to the equilibrium regime. The aging regime corresponds instead to the case in which the two contributions are of the same order {\it i.e.} $|t_1-t_3|=O(t_1)$ or equivalenty to the limit in which we send $t_1$ to infinity while keeping  $a=t_3/t_1$ finite. In this limit both $C(t_1,t_3)$ and $R(t_1,t_3)$ go to zero and we are naturally led to the following ansatz: 
\beqa
C(t,a \, t) &  = & {1 \over t^{\nu}} \CC(a)
\\
R(t,a \, t) & = &  {1 \over t^{\nu+1}} \RR(a)
\eeqa
The scaling exponents $1/t^{\nu}$ and $1/t^{\nu+1}$ are fixed by the matching with the equilibrium behaviour which is obtained for $a\rightarrow 1$.
Plugging the above ansatz into equations (\ref{CORRE}) and (\ref{RESPO}) we obtain the two quadratic equations already presented in the introduction:
\begin{widetext}
\beq
\label{eq_scal_c}
\begin{split}
&\int_0^a \, a^{-\nu} \RR(b) \CC\left( \frac{b}{a} \right) \, db + \int_0^a \left[  \CC(b) a^{-\nu -1} \RR \left(\frac{b}{a}\right) - \CC(a) a^{-\nu -1} \RR_{eq} \left( \frac{b}{a} \right) \right] \, db - \\
& -  \CC(a) a^{-\nu} \CC_{eq}(0) - \CC(a) \CC_{eq}(a) + \int_a^1 \left[  \RR(b) b^{-\nu} \CC \left(\frac{a}{b}\right) - \RR_{eq}(b) \CC(a)\right] \, db + \l \CC^2(a)=0
\end{split}
\eeq
and
\beq
\begin{split}
\label{eq_scal_r}
& \int_a^1 \, \left\lbrace \RR(b) b^{-\nu -1} \RR \left( \frac{a}{b} \right) - \RR(a) \left[ \RR_{eq}(b) +  \RR_{eq} \left( \frac{a}{b} \right)\right] \right\rbrace \, db - \\
& - 2 \RR(a)\CC_{eq}(a) + 2\l \RR(a) \CC(a)=0
\end{split}
\eeq
\end{widetext}
These equations generalize to the off-equilibrium case the critical equilibrium equations that correspond to the last line of Eq. (\ref{CORRE}). Therefore the universal scaling function $\RR(a)$ and $\CC(a)$ are determined (up to a model dependent constant) solely by the parameter $\lambda=w_2/w_1$. The model-dependent constant cannot be determined in this framework and have to be fixed through a matching with the equilibrium soluition at small time differences.
For $\lambda=0$ one can check that (\ref{solf1}) and (\ref{solf2}) provide a solution of the above equations. 
We note that the above equations reduce in the limit $a \rightarrow 1$ to the equilibrium case in which the simple relationship between $\lambda$ and $\nu$ can be obtained. Unfortunately, it seems that such a simplification does not occur for the exponent $\alpha$ controlling the small $a$ behaviour: its determination requires the complete solution of the universal equations. 
In the next section we will introduce a numerical method to solve the equations. 
As discussed in the section \ref{sec:scenario} the critical behaviour of the energy is controlled at leading order by the quadratic term in the action:
\beq
\int \delta Q(1,2)^2 d1 d2
\eeq
Note that the double integration however makes this term vanish \footnote{This result is obvious at equilibrium, see \cite{w1w2}, but remains  true also off-equilibrium because it is just a consequence of causality}, this corresponds to what happens in the Replica method due to the $n\rightarrow 0$ limit because the above term evaluate to $n(n-1)q^2$.
In order to measure the energy at a given time $t_1$ one must consider a fluctuation of the temperature at that given time, therefore breaking the time-traslational invariance of the Hamiltonian, from this it follows that:
\beq
\begin{split}
E(t_1) \propto & \int \delta Q(1,2)^2 d2\\
&=\int_0^{t_1} C(t_1,s)R(t_1,s)ds-C_{eq}^2(0) 
\end{split}
\eeq
the above expression can be simplified using $C^2_{eq}(0)=
\int_{-\infty}^{t_1^+} C_{eq}(t_1,t_2) R_{eq}(t_1,t_2)dt_2$ and leads to the result quoted in section \ref{sec:scenario}:
\beq
E(t)=E_{\infty}+{a_E(\nu) \over t^{2 \nu}}
\eeq
where $E_\infty$ is the model-depedent equilibrium value of the energy and the constant $a_E(\nu)$ is given by:
\beq
\begin{split}
&a_E(\nu)=a_T \, c^2 \left[ \int_0^1 \left[ \CC(a)\RR(a)-\CC_{eq}(a)\RR_{eq}(a) \right] -{1 \over \nu^2}  \right] 
\end{split}
\eeq

\section{Variational solution of the universal equations}
\label{sec:numsol}
We want to obtain the shape of the scaling functions $\CC(a)$ and $\RR(a)$ and, in particular, their power law behaviour 
in $a \simeq 0$ which, through the matching arguments described in Sec. \ref{sec:general}, determines the decay exponent 
of the remanent magnetization. \\ 
In order to solve the equations we use a variational method with an objective function that is simply 
the sum of Eq.s (\ref{eq_scal_c}) and (\ref{eq_scal_r}) squared computed in a set $\Omega$ of $k$ points $\Omega=\{ a_1, \cdots , a_k \}$.\\
Clearly we cannot perform the minimization of the objective function in the entire space of functions $\CC$ and $\RR$ defined on the 
interval $[0,1]$ and we have to choose a trial form. A quite natural choice is the following
\beq
\label{trialC}
\CC(a)= \frac{2 a^{\a}}{\nu (1+a) (1-a)^{\nu}} \left[ 1+ \sum_{i=1}^{O} \CC_i (a-1)^i \right] 
\eeq

\beq
\label{trialR}
\RR(a)= \frac{ a^{\a- 1}}{(1-a)^{\nu +1}} \left[ 1+ \sum_{i=1}^{O} \RR_i (a-1)^i \right] 
\eeq
where we take the form of the scaling functions for $\lambda=0$ and multiply it by a polynomial correction of order $O$. 

\begin{figure}
\includegraphics[width=.99\columnwidth]{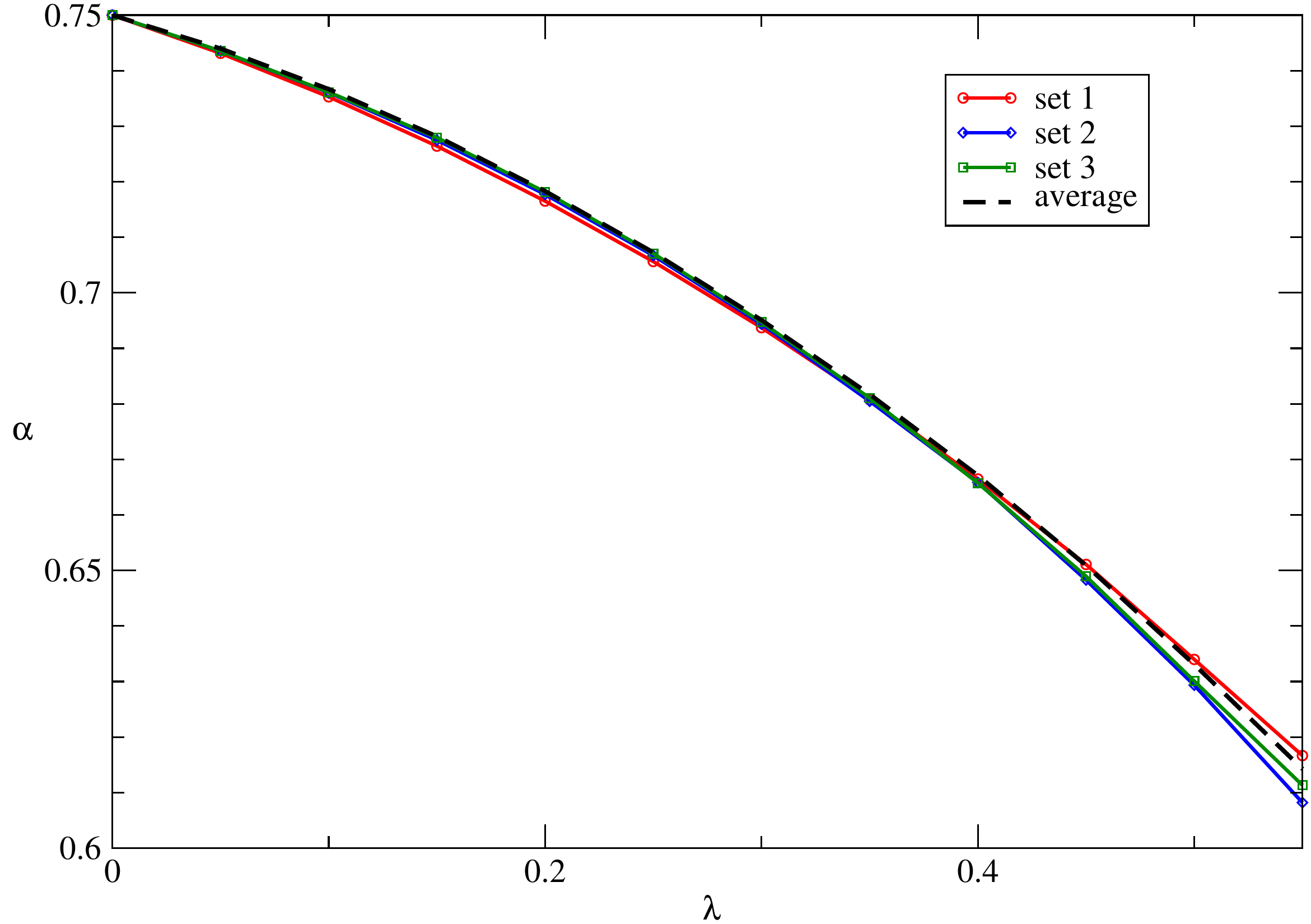}
\caption{(Color online) The exponent $\a$ for three representative choices of $\Omega$. The dashed 
line is the average over these three sets $\Omega_1$, $\Omega_2$ and $\Omega_3$.}
\label{fig:esp}
\end{figure}

\begin{figure}
\includegraphics[width=.99\columnwidth]{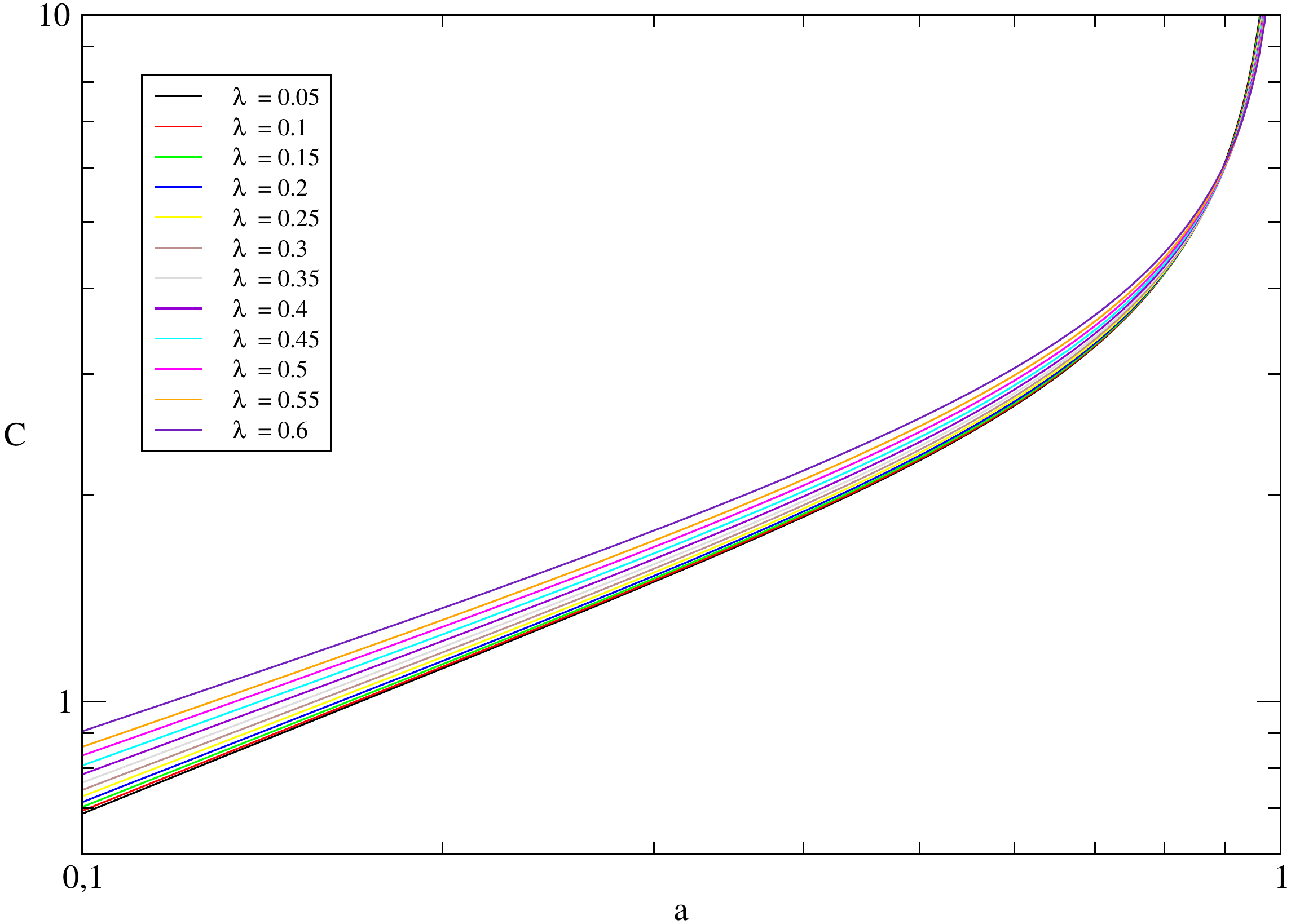}
\caption{(Color online) Correlation scaling function 
for different values of the exponent parameter $\l$ (growing from lower curve to upper curve) and for the particular choice $\Omega=\Omega_1$.}
\label{fig:logcorr}
\end{figure}

\begin{figure}
\includegraphics[width=.99\columnwidth]{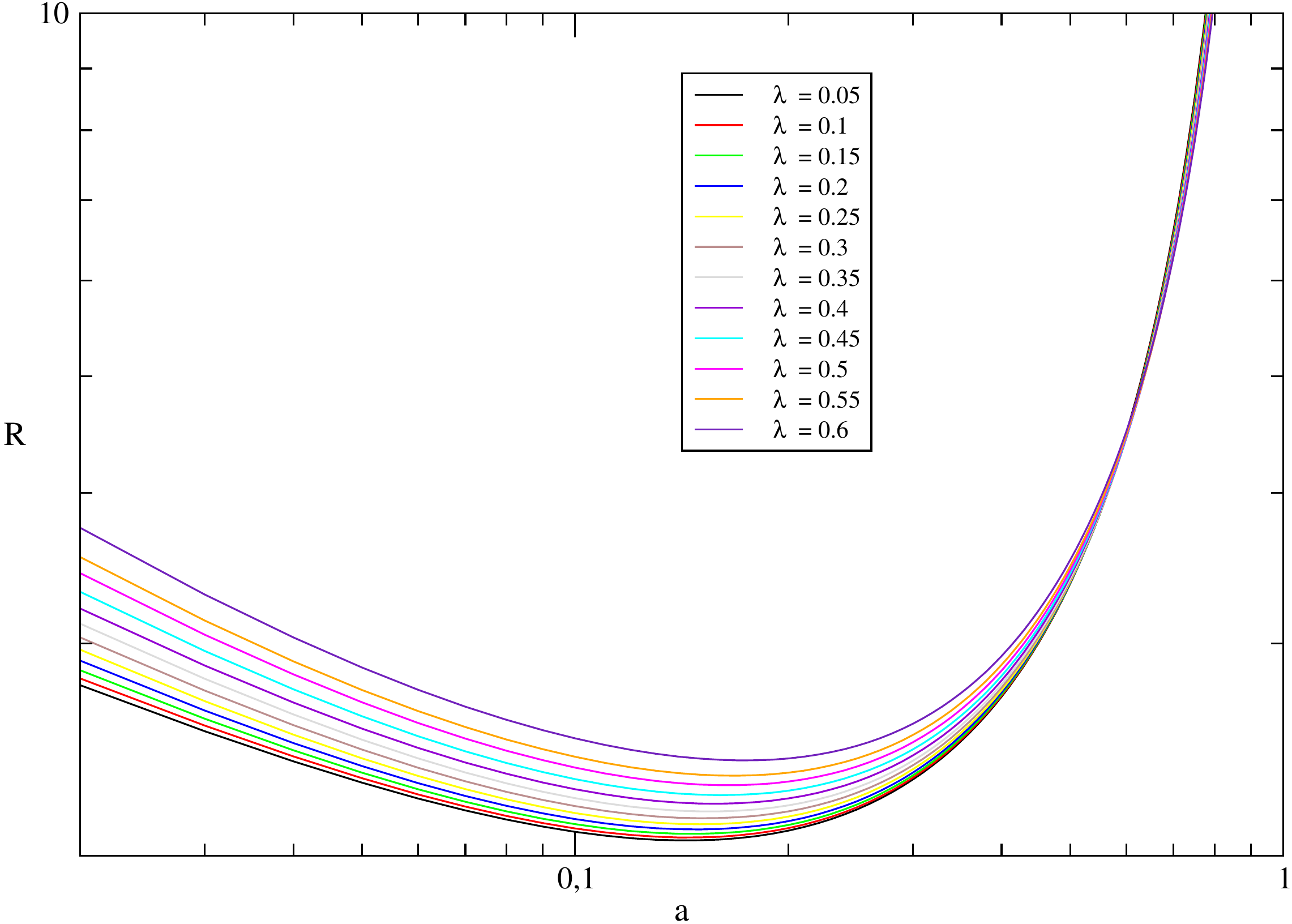}
\caption{(Color online) Response scaling function 
for different values of the exponent parameter $\l$ (growing from lower curve to upper curve) and for the particular choice $\Omega=\Omega_1$.}
\label{fig:logresp}
\end{figure}

In the present case this minimization procedure will determine the optimal value of $\a$ (that is the most relevant quantity) and of the parameters $\{ \CC_i \}$ and $\{ \RR_i \}$. We recall that the value of the equilibrium exponent $\nu$ is known analytically from static computations \cite{w1w2}. \\
Two observations are in order at this point, based on the asymptotic analysis given in Appendix \ref{app:asymptotics}: 
\begin{itemize}
\item for $\l \neq 0$ the first subleading correction to the behaviour of the correlation and response in $a=0$  must necessarily be non-analytic, in the sense that it is some non-integer positive power of $a$ that cannot be expressed as a power series. Despite this, we expect (and verify) that with 
our choice of the trial functions (\ref{trialC}) and (\ref{trialR}) we are able to determine accurately the leading behaviour 
of $\CC$ and $\RR$ for small $a$, that is given by the exponent $\a$. 
\item with our choice of the trial functions the equations have a singular behaviour in $a \sim 0$ and $a \sim 1$. For this reason, 
the equations must be properly re-weighted in the objective function to ensure that both the equations in all the points of $\Omega$ have approximately the same 
relevance in the minimization procedure. 
\end{itemize}
If we call $\EE_C [\a,\{\CC \},\{\RR \}]$ and $\EE_R [\a,\{\CC \},\{\RR \}]$ respectively the l.h.s. of Eq.s (\ref{eq_scal_c}) and (\ref{eq_scal_r}), we show 
in Appendix \ref{app:asymptotics} that they behave as
\beq
\begin{split}
&\EE_C [\a,\{\CC \},\{\RR \}] \propto a^{\a -\nu}    \,\,\,\,\,\,\,\,\,\,\,\,\,\,\,\,\,\,\,\,\,\, \text{for} \,\,\,\,\, a\simeq 0  \\
&\EE_R [\a,\{\CC \},\{\RR \}] \propto a^{\a -\nu - 1}
\end{split}
\eeq
and
\beq
\begin{split}
&\EE_C [\a,\{\CC \},\{\RR \}] \propto (1-a)^{ -2\nu}  \,\,\,\,\,\,\,\,\,\,\,\,\, \text{for} \,\,\,\,\, a\simeq 1  \\
&\EE_R [\a,\{\CC \},\{\RR \}] \propto  (1-a)^{ -2\nu - 1} 
\end{split}
\eeq
All these observations lead to the following form for the objective function
\beq
\label{functional}
\begin{split}
\FF[\a, & \{\CC \},\{\RR \}]=\\
&\sum_{a \in \Omega} \left\lbrace \Bigl(\WW_C(a) \, \EE_C[\a,\{\CC \},\{\RR \}]\Bigr)^2 \right.\\
&+ \left. \Bigl(\WW_R(a) \, \EE_R[\a,\{\CC \},\{\RR \}]\Bigr)^2 \right\rbrace
\end{split}
\eeq
with
\beq
\begin{split}
&\WW_C(a)=a^{\nu - \ol{\a}}(1-a)^{2\nu}\\
&\WW_R(a)=a^{\nu-\ol{\a}+1}(1-a)^{2\nu +1}
\end{split}
\eeq
We have minimized the objective function by means of the Gauss-Newton algorithm which is standard for least squares functions.
Note that the $\ol{\a}$ in (\ref{functional}) should be the correct $\a$ which 
we actually determine with the minimization of  (\ref{functional}) itself. This issue is solved starting with a trial 
value of $\ol{\a}$ and adjusting it self-consistently at each step of the Gauss-Newton algorithm with the value 
at the immediately preceding step. \\ 
We applied the Gauss-Newton algorithm for values of $\l$ up to $0.55$. 
For the trial function we choose $\OO=6$ since, for polynomials of higher orders, 
the convergence of the minimization algorithm becomes quite slow, especially for large $\l$. In any case, we 
observe that for low enough values of $\l$ there is no significant difference in the determination of the exponent 
$\a$ between the case $\OO=6$ and $\OO=8$.
The choice of the set of points $\Omega$ is important for two different reasons: 
\begin{itemize}
\item the number of points must be grater than $\OO$, otherwise 
		the objective function will have flat directions and the Gauss-Newton algorithm 
		will not converge
\item since the trial function is only approximate, the choice of the set of points 
		influences the final result. This dependence on $\Omega$ becomes stronger 
		for larger values of $\l$ while it is almost irrelevant for small $\l$. 
\end{itemize}
Another technical point is that the computation of the objective function requires the numerical evaluation of various definite integrals with arguments that are singular at the extrema of integration. Therefore in order to reduce the numerical errors it is convenient to eliminate the singularities analytically (trough integration by parts) before performing the actual numerical integration. 
We perform the minimization for different $\Omega$s, then we take the average over 
the choices as the correct result and the square root of the variance as our error. 
The results are shown in Tab. \ref{tab:results}. In Fig \ref{fig:esp} we reported the 
exponent $\a$ for three representative choices of the set of points, in particular
\beq
\begin{split}
\Omega_1 =\{ &0.01,0.1,0.2,0.3,0.4,0.5,0.6,0.7,0.8,0.9,0.95 \} \\
\Omega_2 =\{ &0.01,0.05,0.15,0.25,0.35,0.45,0.55,0.65, \\
&0.75,0.85,0.95 \} \\
\Omega_3 =\{ &0.001,0.005,0.01,0.03,0.07,0.15,0.2,0.25, \\ 
&0.35,0.45,0.60,0.75,0.85,0.95 \} \\
\end{split}
\eeq
In Figs. \ref{fig:logcorr} and \ref{fig:logresp} we show the correlation and response scaling functions 
for different values of the exponent parameter $\l$ and for the particular choice $\Omega=\Omega_1$.
The whole procedure was implemented within Mathematica using the routine $\mathrm NIntegrate[]$ for numerical integrations.

\begin{table}
\centering
  \begin{tabular}{| c | c |c|}
    \hline
    $\l$  & $\a$ & \text{Err}\\ \hline
   $0$ & 0.750 & 0\\ \hline
   $0.05$ & 0.744 & 0.002\\ \hline
   $0.1$ & 0.737 & 0.003\\ \hline
   $0.15$ & 0.728 & 0.003\\ \hline
   $0.2$ & 0.718 &  0.003\\ \hline
   $0.25$ & 0.707 & 0.003\\ \hline
   $0.3$ & 0.695 & 0.003\\ \hline
   $0.35$ & 0.682 & 0.003\\ \hline
   $0.4$ & 0.667 & 0.004\\ \hline
   $0.45$ & 0.651 & 0.005\\ \hline
   $0.5$ & 0.633 & 0.007\\ \hline
   $0.55$ & 0.614 &  0.009\\ \hline
\end{tabular}
\caption{The exponent $\a$ determined through the Gauss-Newton minimization procedure. 
The value for $\l=0$ has zero error since it is determined analytically. The other values reported in the 
table are averages over different choices for the set of points $\Omega$ with the associated error.}
\label{tab:results}
\end{table}

\section{Tests}
\label{sec:valther}
In this section we present two validations of the theory presented above: the first one is a Monte Carlo study of the three colors 
fully-connected Potts model, while the second is a power series solution of the dynamical equations for the spherical $(2+3)$-spin model.
\subsection{Monte Carlo study of the $3$-colors fully-connected Potts model}
We consider the fully-connected $3$-colors Potts Hamiltonian 
\beq
\HH = - \frac{1}{2}\sum_{i\neq j} J_{ij} \left( 3 \, \d_{\s_i ,\s_j} - 1 \right)
\eeq
where the couplings are i.i.d. Gaussian random variables with zero mean and 
variance $1/N$. \\
This system undergoes a continuous transition at the critical temperature $T_c=1$ with $\l = 1/2$ \cite{Gross85,Calta12}, 
which gives an equilibrium exponent 
\beq
\nu=0.3953
\eeq 
We study the system by means of an off-equilibrium Monte Carlo simulation starting from a random configuration and, in particular, 
we consider the energy and the remanent magnetization. We simulated fully-connected systems of size $2^9,2^{10},2^{11},2^{12}$ ($1000$ samples) 
and of size $2^{16}$ ($329$ samples).

\begin{figure}[h!]
\includegraphics[width=.99\columnwidth]{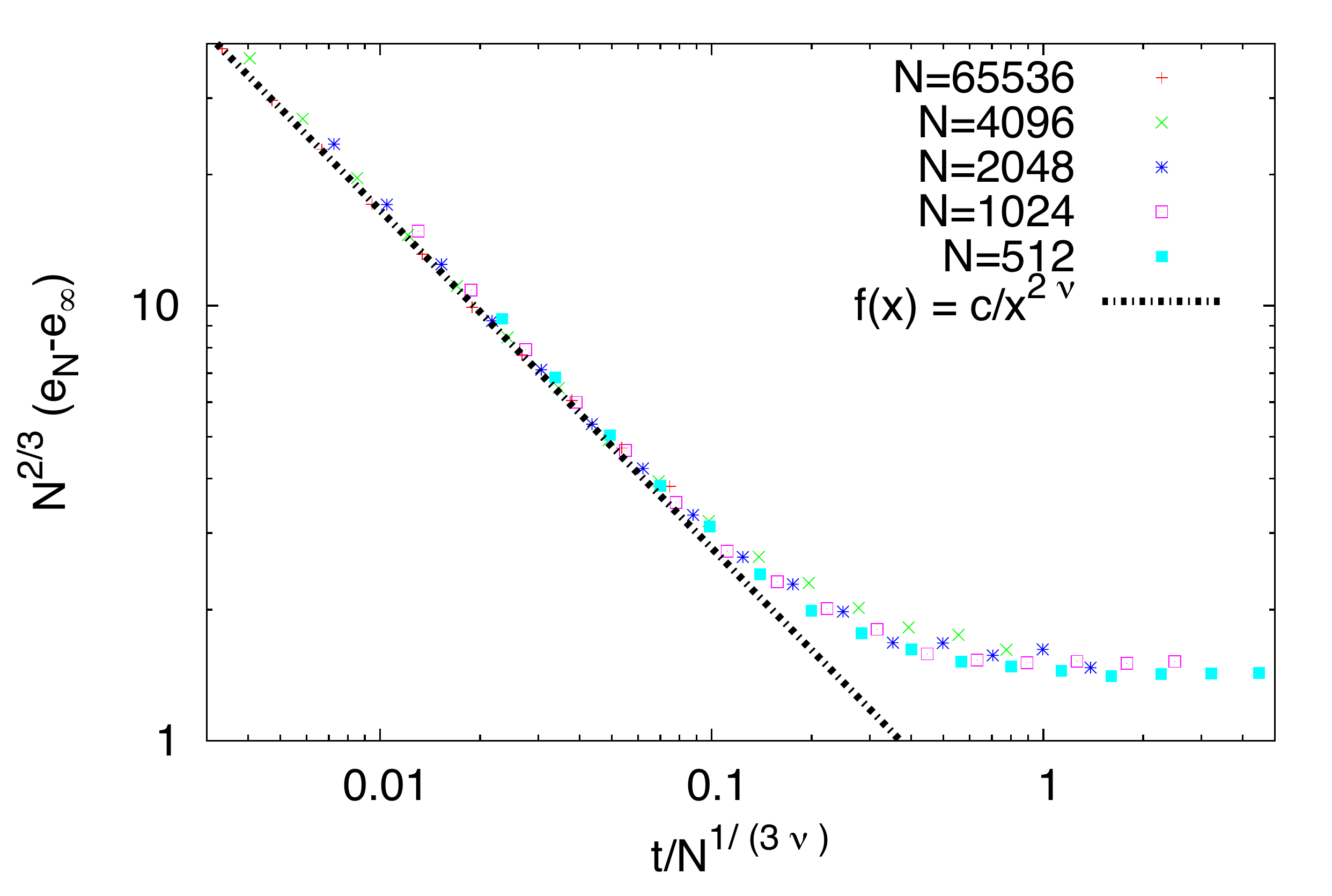}
\caption{Energy decay towards its equilibrium value for different sytstem sizes. The rescaled curves display an excellent collapse, and a nonlinear 
fit with $x^{-\Upsilon}$ gives an exponent $\Upsilon=0.777 \pm 0.010$}
\label{fig:pottsene}
\end{figure}

\begin{figure}[h!]
\includegraphics[width=.99\columnwidth]{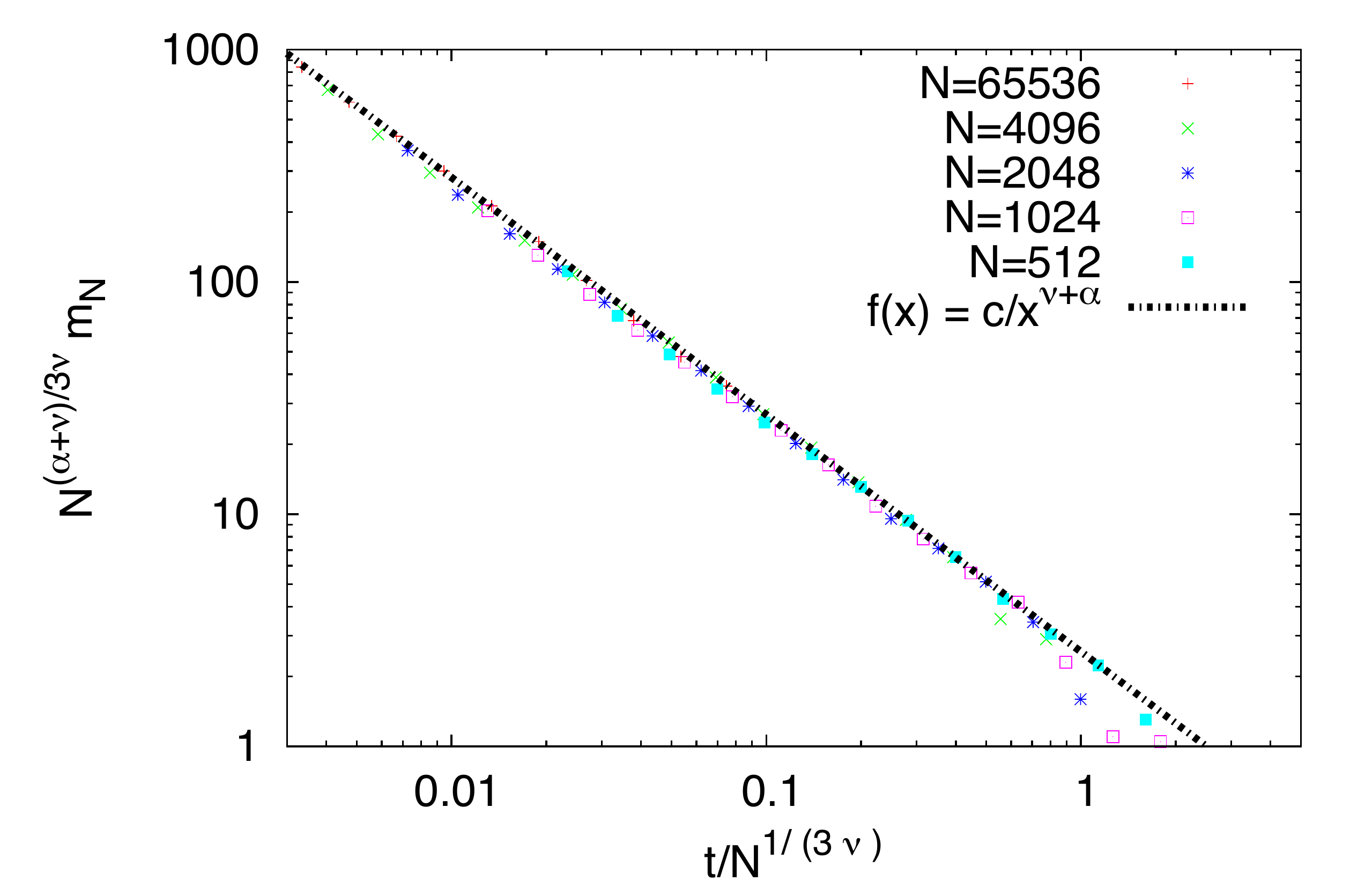}
\caption{Magnetization decay towards zero for different system sizes. The rescaled curves display an excellent collapse, and a nonlinear 
fit with $x^{-\d}$ gives an exponent $\d=1.025 \pm 0.013$}
\label{fig:pottsm}
\end{figure}

Due to finite size effects $e(t)$ and $m_{R}$ display a power law behavior only up to a certain time-scale $t^*(N)$ that 
 diverges with the size as $t^*(N) \simeq N^{1/3\nu}$.
 Moreover, in order to have a collapse of the curves for different sizes we have to take into account the finite-size and finite-time effects, 
 and use the rescaled variables  
\beq
\begin{split}
& t N^{-\frac{1}{3\nu}} \\
&N^{2/3} (e_N-e_{\infty}) \\
&N^{\frac{\a+\nu}{3 \nu}} m_N
\end{split}
\eeq
where $e_N$ and $m_N$ are the energy and magnetization at finite size $N$.
With the rescaled variables we observe an excellent collapse of the energy and magnetization decay, see Figs. \ref{fig:pottsene} and \ref{fig:pottsm}. 

We perform a power-law fit on the curves for the largest size ($2^{16}$) 
assuming for large times 
\beq
\label{curvefit}
\begin{split}
& e(t)-e_{\infty} = \frac{c_e}{t^{\Upsilon}} \\
& m(t)=\frac{c_m}{t^{\d}}
\end{split}
\eeq
and we obtain 
\beq
\begin{split}
&\Upsilon_{mc} = 0.777 \pm 0.010 \\
&\delta_{mc} = 1.025 \pm 0.013
\end{split}
\eeq
where $mc$ stands for Monte Carlo. 
Considering that we know the exact value of the equilibrium exponent $\nu = 0.395 \cdots$, 
we can compute the theoretical value of $\Upsilon$, that is $\Upsilon_{th} = 0.790$. 
If compared with the Monte Carlo estimate we can see that the agreement is good (within 2$\s$). 

Now, in order to give our Monte Carlo estimate of $\a$ we have two options: 
\beq
\a^{(1)}_{mc} = \d_{mc} - \Upsilon_{mc}/2 = 0.633 \pm 0.018
\eeq
or
\beq
\a^{(2)}_{mc} = \d_{mc} - \nu = 0.630 \pm 0.013
\eeq

In both the cases there is complete agreement, within the error, with our theoretical estimate $\a_{th}=0.633 \pm 0.007$.

\subsection{Power series solution of the exact equations for the spherical $(2+3)$-spin model}
We consider the Hamiltonian
\beq
\HH=-\sum_{p} \sum_{i_1<i_2<\cdots <i_p} J^{(p)}_{i_1 \cdots i_p} \,  s_{i_1} \cdots s_{i_p}
\label{HH}
\eeq
where the $s_i$ are continuous spins subject to a global spherical constraint 
\beq
\sum_{i=1}^{N} s_i^2=N
\eeq
and the couplings are uncorrelated Gaussian variables with zero mean and variance 
\beq
\ol{(J^{(p)})^{2}}=\frac{J_p^2 p!}{2N^{p-1}}
\eeq

If we define $\mu_p = J_p^2$ and the function

\beq
\Phi(x)=\sum_p \mu_p \, x^{p}
\eeq

the dynamical equations can be written in the following way

\begin{widetext}

\begin{eqnarray}
\nonumber
\mu(t)&=&T+\frac{1}{2} \int_0^t \, ds \, R(t,s) \Phi '\left[ C(t,s)\right] + \frac{1}{2}\int_0^t \, ds \,C(t,s) R(t,s) \Phi ''\left[ C(t,s)\right]\\
\frac{\partial R(t,t')}{\partial t}&=&-\mu(t)R(t,t')+ \frac{1}{2} \int_{t'}^{t} \, ds \, \Phi ''\left[ C(t,s)\right] R(t,s) R(s,t')\\
\nonumber
\frac{\partial C(t,t')}{\partial t}&=& -\mu(t)C(t,t') + \frac{1}{2} \int_0^{t'} \, ds \, \Phi '\left[ C(t,s)\right] R(t',s) + \frac{1}{2} \int_0^{t} \, ds \, \Phi ''\left[ C(t,s)\right] R(t,s) C(s,t') 
\end{eqnarray}

We are interested in the specific case of the spherical $(2+3)$-spin model where the equations become straightforwardly

\begin{eqnarray}
\nonumber
\mu(t)&=&T+2\mu_2 \int_0^t \, ds \, R(t,s) C(t,s) + \frac{9}{2} \mu_3 \int_0^t \, ds \, R(t,s) C^2(t,s)\\
\label{eq_2piu3}
\frac{\partial R(t,t')}{\partial t}&=&-\mu(t)R(t,t')+ \mu_2 \int_{t'}^{t} \, ds \, R(t,s) R(s,t') + 3\mu_3 \int_{t'}^{t} \, ds \, C(t,s) R(t,s) R(s,t')\\
\nonumber
\frac{\partial C(t,t')}{\partial t}&=& -\mu(t)C(t,t') + \mu_2 \int_0^{t'} \, ds \,C(t,s) R(t',s)  + \frac{3}{2} \mu_3 \int_0^{t'} \, ds \,C^2(t,s) R(t',s) \\
\nonumber
 &+& \mu_2 \int_0^{t} \, ds \, R(t,s) C(s,t')  + 3\mu_3 \int_0^{t} \, ds \, C(t,s) R(t,s) C(s,t') 
\end{eqnarray}

\end{widetext}

\begin{figure}
\includegraphics[width=.99\columnwidth]{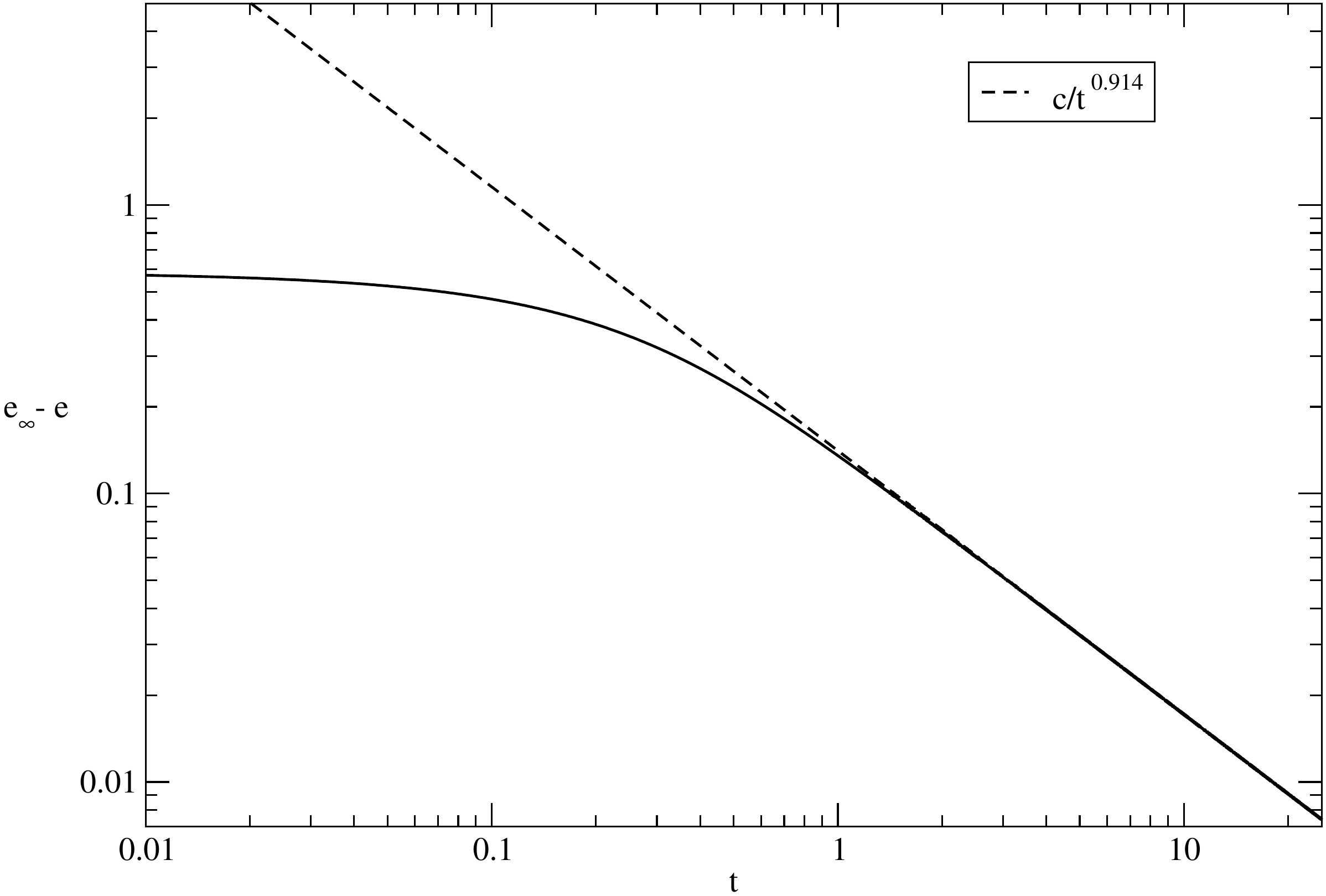}
\caption{Energy decay from the power series solution of the exact equation for the spherical $(2+3)$-spin model. The nonlinear 
fit gives an exponent $\Upsilon_{ps}=0.914$.}
\label{fig:multiplier}
\end{figure}

\begin{figure}
\includegraphics[width=.99\columnwidth]{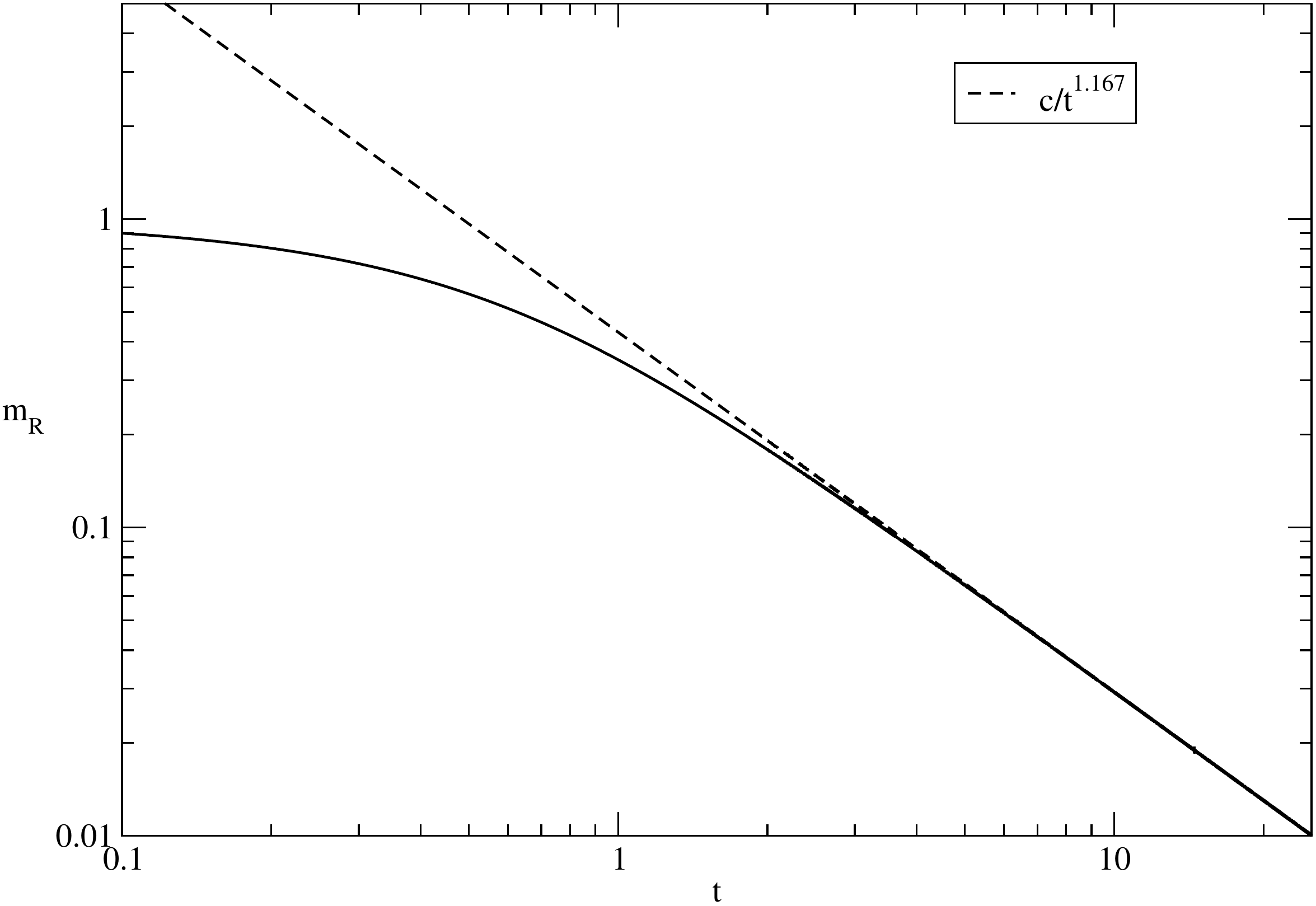}
\caption{Magnetization decay from the power series solution of the exact equation for the spherical $(2+3)$-spin model. The nonlinear 
fit gives an exponent $\delta_{ps}=1.167$.}
\label{fig:mtrm}
\end{figure}

In this case, as well as in the case of the spherical $2$-spin model, we can see that $C(t,0)=R(t,0)$ since they satisfy the very same equation, namely
\beq
\begin{split}
\frac{\partial R(t,0)}{\partial t} & = -\mu(t)R(t,0) \\
&+ \mu_2 \int_{0}^{t} \, ds \, R(t,s) R(s,0) \\
&+ 3\mu_3 \int_{0}^{t} \, ds \, R(t,s)^2 R(s,0)
\end{split}
\eeq
where we have already considered the equality of correlation and response.\\
\indent We are able to solve Eq.s (\ref{eq_2piu3}) in power series of the two times $t$ and $t'$ starting at $t_0=0$. 
The resulting (truncated) asymptotic series can be resummed using Pad\'e approximants. 
It can be shown easily from static computations \cite{Crisanti04b,Crisanti06} that the above model corresponds to our 
universal equations (\ref{eq_scal_c}) and (\ref{eq_scal_r}) with 
\beq
\l=\frac{3\mu_3}{2\mu_2}
\eeq
In particular we choose $\mu_2=1$ and $\mu_3=1/6$ yielding
\beq
\l=\frac{1}{4}
\eeq
We computed the series to 163 orders and resummed it with Pad\'e approximants of order $(80,80)$, the results are shown 
in Figs. \ref{fig:multiplier} and \ref{fig:mtrm} respectively for the energy and for the remanent magnetization.
In this case we are not actually able to determine 
an error on the measure of the exponents since our points are exact. The source of the error in the determination of $\nu$ and 
$\a$ is only the fact that with the resummed series we are not able to converge at very large times. For this reason we may be 
quite far from the true asymptotic power-law regime. Despite this fact, as we will see, the results are in reasonably good agreement 
with our predictions.

Fitting the results with a power law for $t \in [10,20]$ we get 
\beq
\Upsilon_{ps} = 0.914\pm 0.010
\eeq
and
\beq
\delta_{ps} = 1.167 \pm 0.010
\eeq

where the error is roughly estimated considering that, choosing different 
time intrvals for the fit, we get slightly different results. 
The exact value of the equilibrium exponent is $\nu=.455073$ and 
the estimate from the power series solution is $\nu_{ps} = 0.457 \pm 0.005$. 

As in the preceding case, in order to give our power series estimate of $\a$ we have two options: 
\beq
\a^{(1)}_{ps} = \d_{ps} - \Upsilon_{ps}/2 = 0.710 \pm 0.015
\eeq
or
\beq
\a^{(2)}_{ps} = \d_{ps} - \nu = 0.712 \pm 0.010
\eeq

we predict from our theoretical analysis $\a_{th}=0.707\pm 0.003$. As already said, these values are in good agreement 
with the results from the series expansion despite the difficulty of the measure. 

In the introduction we pointed out that action (\ref{repgibb}) and all the results we derived from it can be applied provided the Hamiltonian possess some additional symmetries, {\it e.g.} time-reversal in magnetic systems. From this it follows that in general a Hamiltonian like (\ref{HH}) with non-vanishing odd-$p$ terms {\it cannot} be mapped into action (\ref{repgibb}). In the present section we could successfully apply the theory to the $2+3$ case because the model is defined on a fully-connected lattice and the effect of breaking time reversal  vanishes in the thermodynamic limit. For the same reason it follows that the present theory applies also to the same models with odd-$p$ interactions defined on random lattices, but not on lattices in finite dimension.

\section{Conclusions}
\label{sec:conclusion}

We have formulated a general scenario for the off-equilibrium critical behavior of a class of glassy systems characterized by a specific structure of the replicated Gibbs free energy.

The off-equilibrium correlation and response functions obey a precise scaling form in the aging regime. The structure of the {\it equilibrium} replicated Gibbs free energy fixes the corresponding {\it off-equilibrium} scaling functions implicitly through two functional equations.
The details of the model enter these equations only through the ratio $w_2/w_1$  of the cubic coefficients (proper vertexes) of the replicated Gibbs free energy. Therefore the scaling functions and exponents are controlled  by the very same parameter exponent $\lambda=w_2/w_1$ that determines equilibrium dynamics according to \cite{w1w2}.

The dynamical exponent $\Upsilon$ describing the approach to equilibrium of the energy turns out to be $\Upsilon=2 \nu$ where $\nu$ is the dynamical exponent of the decay of the equilibrium correlation at criticality that obeys the well-known relationship $\lambda=\Gamma^2(1-\nu)/\Gamma(1-2\nu)$. The dynamical exponent associated to the decay of the remanent  magnetization is determined by $\delta=\alpha+\nu$ where $\alpha$ is the exponent associated to the behavior of the scaling functions at small arguments.

The off-equilibrium universal equations are a generalization of the scale-invariant equations obtained at equilibrium. We have exhibited the analytical solution for $\lambda=0$ but no analytical solution is known for general values of $\lambda$.  Finding approximate solutions is not at all trivial because the scaling functions are singular at the extrema.   
Nevertheless we have devised an approximation scheme that appears to yield consistent results at least for not too large values of $\lambda$. The theory have been validated by means of i) exact analytical computation in the spherical SK model, ii) large-time analytical computation in the SK model, iii) numerical simulations in the three-states Potts glass and iv) small-time power-series solution of the full dynamical equations in multi-$p$-spin spherical models that also correspond to some schematic MCT models. 

In summary, the main result of the present paper is that the equilibrium replicated Gibbs free energy determines off-equilibrium critical dynamics at large times both qualitatively (through its structure) and quantitatively (through the actual value of the cubic proper vertexes). We expect that similar results hold for other types of transition as well. In particular it would be interesting to extend this analysis to the continuous SG transition in a field and to the discontinuous SG transition, that corresponds within MCT to the standard liquid-glass transition (generic $A_2$ singularity).

\subsection*{Acknowledgements}

We thank Alain Billoire for discussions.
The research leading to these results has received financial support 
from the European Research Council (ERC)
grant agreement no [247328].

\appendix

\section{Critical approach to equilibrium in the SK model at the critical temperature.}
\label{app:sk}

The Hamiltonian reads
\begin{equation}
H[\sigma]=-\sum_{i,k} J_{i,k}\sigma_{i}\sigma_{k} \ .
\end{equation}
with $\s_i = \pm 1$.
In order to write down exact equations we avoid to consider Monte Carlo dynamics or continuous time dynamics: exact dynamical equations can be written, but they are not so simple. We consider a generalized model where the Hamiltonian is given by:

\begin{equation}
H_{P}[\sigma,\tau]=-\sum_{i,k} J_{i,k}\sigma_{i}\tau_{k} \ .
\end{equation}
It has been shown that (as far as the free energy is concerned) this Hamiltonian has the same equilibrium properties of the usual SK Hamiltonian, where $\tau_{i}=\sigma_{i}$. A sequential update of this Hamiltonian corresponds to two steps of parallel update in the SK model, where the $\sigma$ are the spin at even time and the $\tau$ are the spins at odd time.

The dynamics that we are considering is therefore parallel update of the spins for the SK model using a local heath bath dynamics, i.e the spins at time $(t+1)$ have a probability distribution given by 
\begin{equation}
P_{t+1}[\sigma(t+1)]\propto \exp \left(\beta\sum_{i,k} J_{i,k}\sigma_{i}(t+1)\sigma_{k}(t)\right)\ .
\end{equation}

With this equation of motion, we can write exact recursion equations. For example for the magnetization we have
\begin{equation}
m_{i}(t+1)=\tanh\left(\beta\sum_{k} J_{i,k}\sigma_{k}(t)\right) \label{KEY}\ .
\end{equation}

In order to compute the time evolution we write
\begin{equation}
P_{t}[\sigma]\propto \exp \left(-\beta H_{t}[\sigma]\right)
\end{equation}
We have that 
\begin{equation}
H_{t}[\sigma]=H[\sigma]+\Delta H_{t}[\sigma]
\end{equation}
where for large time $\Delta H_{t}[\sigma]$ must go to zero.

We suppose that for large times $\Delta H_{t}[\sigma]$ has a simple expression: i.e.
\begin{equation}
\Delta H_{t}[\sigma]=\sum_{i}h_{i}(t)\sigma_{i}.
\end{equation}
Two (and more) spins interactions are assumed to give higher order corrections. 
The consistency of this approximation can be checked by considering the perturbative effect of a possible term in $\Delta H_{t}[\sigma]$ proportional to $\sum_{i,k} R_{i,k}(t)\sigma_{i}\sigma_{k}$.

If we stay in the situation where there is no replica symmetry breaking for the Hamiltonian $H_{t}$, we have that the computation of the r.h.s. of  equation (\ref{KEY}) is easy and the computation is the same of the one of the cavity approximation. We finally get
\begin{equation}
\begin{split}
m_{i}(t+1)&=\tanh\Big(\beta\sum_{k} J_{i,k}m_{k}(t)  \\
&- \beta^{2}(1-q(t))m_{i}(t)\Big) 
\end{split}
\label{FINAL}\, ,
\end{equation}
where 
\begin{equation}
q(t)=1/N \sum_{i}m_{i}(t)^{2}
\end{equation}

At large times, where $m_{i}(t+1)$ has a smooth dependence on the integer value time $t$, we can use for simplicity a continuous time $t$ and write
\begin{equation}
\begin{split}
{d\, m_{i}(t)\over dt}&=\tanh\Big(\beta\sum_{k} J_{i,k}m_{k}(t) \\
&-\beta^{2}(1-q(t))m_{i}(t)\Big)-m_{i}(t)
\end{split}
\label{CON}.
\end{equation}

We can now use the spectral properties of the matrix following \cite{Ranieri97} with similar results.

We study the problem at the critical point where $\beta=1$,

\begin{equation}
{d \, m_{\lambda}(t) \over d t}=-(\lambda+q(t)) m_{\lambda}(t)
\end{equation}
where $m_{\lambda}(t)$ is the projection of the magnetization on the eigenvector of $J$ with eigenvalue $2-\lambda$ ($0\le \lambda \le 4$).
Consequently we have 
\begin{equation}
m_{\lambda}(t) \propto r_{\lambda}\exp(-\beta \lambda t +I(t))
\end{equation}
where $I(t)=\int_{1}^{t}q(t)$.

Now if $q(t)$ behaves as 
\begin{equation}
q(t)=\alpha/t  + o \left( t^{-1} \right) 
\end{equation}
we have $I(t)\propto t^{\alpha}$. On the other hand
\begin{equation}
\begin{split}
q(t)&=\int d\mu(\lambda)m_{\lambda}(t)^{2} \\
&\propto\int_{0}^{4}d\lambda \sqrt{\lambda}t^{2\alpha} \exp(-\beta \lambda t)\propto t^{-3/2+2 \alpha}
\end{split}
\end{equation}
The only consistent solution is $\alpha=1/4$.

One finds that 
\begin{equation}
m_{\lambda}(t) \propto r_{\lambda}t^{1/4}\exp(-\beta \lambda t) \, ,
\end{equation}

Finally $m_{r}(t)\propto t^{-5/4}$ and $q(t)\propto t^{-1}$

Now we want to study the behaviour of the energy and 
we should be precise with the definitions. Two are possible choices:
\begin{itemize} 
 \item[(a)] $H[\sigma(t)]$  
 \item[(b)] $H_{P}[\sigma(t),\sigma(t)]$
\end{itemize}
We consider here case (a). Here we have to compute $1/2\langle \sigma_{i}(t)\sigma_{k}(t) J_{i,k}\rangle_{t}$. The computation should be done with some care, because there is a small correlation between the two spins, that is given by $\beta J_{i,k}\langle\sigma_{i}^{2}\rangle_{t}^{c}\langle\sigma_{k}^{2}\rangle_{t}^{c}$, where $\langle ...\rangle^{c}$ is the connected expectation value. We finally obtain

\begin{equation}
N E(t)= \frac12 \langle \sigma_{i}\sigma_{k}\rangle_{t} J_{i,k}- N \frac12 \beta (1-q(t))^{2}
\end{equation}
while the first term decays as $1/t^{2}$ the second term gives the leading contribution to 
\begin{equation}
E(t)=-1/2+q(t)\approx -1/2+\alpha t^{-1}
\end{equation}
The same results are obtained considering definition (b).

\section{Solution of the spherical $2$-spin model}
\label{app:spherical}
The Langevin equation for the spherical $2$-spin model can be projected on the eigenvalues 
of the interaction and reads \cite{dedom_giardina}
\beq
\frac{\d s_{\mu}(t)}{\d t}=[\mu - z(t)]s_{\mu}(t) + \eta_{\mu}(t)
\eeq
where $\mu \in [-2J,2J]$ is an eigenvalue of the interaction matrix, $z(t)$ is the Lagrange multiplier enforcing the spherical constraint 
and $\eta$ is a Gaussian noise with 
\beq
\begin{split}
&\langle \eta_{\mu} (t) \rangle =0\\
&\langle \eta_{\mu}(t) \eta_{\epsilon}(t')\rangle =2T \d_{\mu ,\e} \d(t-t')  
\end{split}
\eeq
Setting the initial time $t_0=0$ and defining 
\beq
\gamma (t)\equiv e^{2Jt-\int_0^t \, ds\, z(s)}
\eeq
the general solution is given by
\beq
\begin{split}
s_{\mu}(t) &= s_{\mu}(0) e^{-(2J-\mu)t} \gamma (t) \\
&+ \int_0^t \, ds \, e^{-(2J-\mu)(t-s) } \eta_{\mu}(s) \frac{\gamma(t)}{\gamma(s)}
\end{split}
\eeq
It can be shown that a random initial condition for the spins $s_i$ corresponds to a fixed uniform initial 
condition $s_{\mu}(0)=1$ for the projections.\\
The spherical constraint with random initial conditions becomes a closed equation for 
\beq
\label{bigD}
D(t)\equiv \frac{1}{\gamma^2(t)}=e^{-4Jt +2\int_0^t \, ds\, z(s)}
\eeq
and the equation is the following 
\beq
\label{constraint}
\begin{split}
D(t)&=\int_{-2J}^{2J} d\mu \, \rho(\mu) \, \left[ \vphantom{\int}  e^{-2(2J-\mu)t } \right. \\
&+ 2T  \left. \int_0^t \, ds \, e^{-2(2J-\mu)(t-s)} D(s) \right]
\end{split}
\eeq
The correlation and the response can be expressed in terms of $\gamma(t)$ once the Lagrange multiplier $z(t)$ is eliminated from the equations. \\
Their form is the following (cf. Eq.s (\ref{corre}) and (\ref{resp}))
\beq
\nonumber
\begin{split}
C(t,t')&=\int_{-2J}^{2J} \, d\mu \, \rho(\mu)  \times \\
&\left[ \vphantom{\int}  s_{\mu}^2(0) e^{-(2J-\mu)(t+t') } \g (t) \g (t') \right. \\ 
&+ \left. 2T \int_0^{\text{min} (t,t')} ds \, e^{-(2J-\mu)(t+t'-2s)} \frac{\g (t) \g (t')}{\g^2(s)}\right]
\end{split}
\eeq
\beq
\nonumber
R(t,t')= \int_{-2J}^{2J} \, d\mu \, \rho(\mu) \, e^{-(2J-\mu)(t-t') } \frac{\g (t)}{\g (t')}
\eeq
Given the above equations, it is clear that once $\gamma(t)$ is determined, the correlation and the response 
can be computed straightforwardly through simple integrations. \\
Taking the Laplace transform Eq. (\ref{constraint}) for the spherical constraint we obtain
\beq
\tilde{D}(u)=\int_{-2J}^{2J} d\mu \, \rho(\mu) \, \frac{1+2T\tilde{D}(u)}{u+2(2J-\mu)} 
\eeq
If we now define
\beq
\tilde{D}(u)=\frac{G(u)}{1-2TG(u)}
\eeq
we obtain $G(u)$ in a closed form
\beq
\begin{split}
G(u)&=\int_{-2J}^{2J} d\mu \, \rho(\mu) \, \frac{1}{u+2(2J-\mu)} \\
&=G(0)-\frac{u}{2} \int_{-2J}^{2J} d\mu \, \rho(\mu) \times \\
&\,\frac{1}{(2J-\mu)(u+2(2J-\mu))}
\end{split}
\eeq
where we added and subtracted 
\beq
\begin{split}
G(0)&=\frac{1}{2T_c} \\
&=\frac{1}{2} \int_{-2J}^{2J} \, \frac{d\mu}{2\pi J} \, \sqrt{4J^2 - \mu ^2} \,\frac{1}{2J-\mu} 
\end{split}
\eeq
We need the leading behavior of $G(u)$ at small $u$ since we are interested in the long-time region of its Laplace anti-transform
\beq
G(u) \simeq \frac{1}{2}\left( \frac{1}{T_c} - cu^{1/2} \right)
\eeq
where $c=(2 J^3)^{-1/2}$. As a consequence, we obtain 
\beq
\tilde{D}(u) \simeq \frac{1}{2} \frac{1-cT_cu^{1/2}}{(T_c-T)+cTT_cu^{1/2}}
\eeq
The leading behavior of $\tilde{D}(u)$ for small $u$ is different if we are {\it at} or {\it below} the critical temperature. 
In Refs. \cite{cugliandolo_dean, dedom_giardina} can be found a detailed treatment of the $T<T_c$ case.\\
At $T=T_c$ we find at leading order 
\beq
\tilde{D}(u)\simeq \frac{1}{(2J)^{3/2}T_c^{2}}u^{-1/2}
\eeq
Taking the inverse transform we obtain
\beq 
D(t)\simeq \frac{1}{(2J)^{3/2}T_c^{2} \pi^{1/2}} t^{-1/2} 
\eeq
Which, setting without loss of generality $J=1$ and, consequently, $T_c=1$ and using definition (\ref{bigD}), yields (cf. Eq. (\ref{gamma}))
\beq
\nonumber
\g (t)\simeq 2^{3/4} \pi^{1/4} t^{1/4}
\eeq

\section{Asymptotic analysis of the universal equations}
\label{app:asymptotics}
We assume that the correlation and response scaling functions $\CC(a)$ and $\RR(a)$ have a power-law 
behaviour in $a \approx 0$, in particular
\beq
\begin{split}
\CC(a)&=a^{\a} \tilde{\CC}(a) \\
\RR(a)&=a^{\b} \tilde{\RR}(a) \\
\end{split}
\eeq
where $\tilde{\CC}(a)$ and $\tilde{\RR}(a)$ are non-singular in $a=0$ and 

\beq
\begin{split}
&\tilde{\CC}(a) \xrightarrow[a\rightarrow 1]{} \CC_{eq}(a)\\
&\tilde{\RR}(a) \xrightarrow[a\rightarrow 1]{} \RR_{eq}(a)
\end{split}
\eeq

We can rephrase Eq. (\ref{eq_scal_c}) in terms of the new {\it tilded} functions obtaining

\beq
\begin{split}
& a^{-\a -\nu} \int_0^{a} b^{\a +\b} \tilde{\RR}(b)  \tilde{\CC}\left( \frac{b}{a} \right) db  \\
+ & a^{ -\nu -\b -1} \int_0^{a}  [b^{\a} \tilde{\CC}(b) - a^{\a} \tilde{\CC}(a)] b^{\b} \tilde{\RR}\left(\frac{b}{a} \right) db  \\
+ & a^{\a -\nu -1} \tilde{\CC}(a) \int_0^{a} \left[  a^{-\b} b^{\b} \tilde{\RR}\left(\frac{b}{a} \right) -  \RR_{eq}\left( \frac{b}{a} \right)\right] db \\
+ & a^{\a} \tilde{\CC}(a) [a^{\a} \tilde{\CC}(a) - a^{-\nu} \CC_{eq}(0) - \CC_{eq}(a)] + \\
+ & a^{\a} \tilde{\CC}(a) \int_{a}^{1} [b^{\b} \tilde{\RR}(b) - \RR_{eq}(b)]  db\\
+ & a^{\a} \int_{a}^{1} b^{\b} \tilde{\RR}(b) \left[b^{-\a -\nu} \tilde{\CC}\left( \frac{a}{b} \right) - \tilde{\CC}(a) \right]  db\\
+ & a^{2 \a} \left( \frac{\omega_2}{\omega_1} - 1  \right) \tilde{\CC}^2(a) = 0
\end{split}
\eeq

We now extract the leading order from each of the terms of the l.h.s. of the above equation

\beq
\label{tildeeq}
\begin{split}
&(a^{\b -\nu +1}) \tilde{\RR}(0) \int_0^1 dy\, y^{\a + \b} \tilde{\CC}(y) + o(a^{\b - \nu +1})\\
+ &(a^{\a -\nu})\tilde{\CC}(0) \int_0^1 dy\, (y^{\a} -1) y^{\b} \tilde{\RR}(y) + o(a^{\a - \nu})\\
+ &(a^{\a -\nu})\tilde{\CC}(0) \int_0^1 dy\,  [y^{\b} \tilde{\RR}(y) - \RR_{eq} (y)] + o(a^{\a - \nu })\\
- &(a^{\a -\nu})\tilde{\CC}(0) \CC_{eq}(0) + o(a^{\a - \nu })\\
- &(a^{\a})\tilde{\CC}(0) \CC_{eq}(0) + o(a^{\a })\\
+ &(a^{\a})\tilde{\CC}(0) \int_0^1 dy\,  [y^{\b} \tilde{\RR}(y) - \RR_{eq} (y)] + o(a^{\a})\\
+ &(a^{\b -\nu +1})\tilde{\RR}(0) \int_0^1 dy\, y^{\a -\b +\nu -2} \tilde{\CC}(y) +o(a^{\b -\nu +1}) \\
+ &(a^{2 \a}) \frac{w_2}{w_1}\tilde{\CC}(0)^2 + o(a^{2\a})= 0
\end{split}
\eeq
Generally speaking, the candidates to be the leading terms in the equation are the 
ones of order $\a - \nu$ and the ones of order $\b - \nu +1$ depending 
on which one is the smallest.\\
\indent For $\l=0$ we know from the exact 
solution of the spherical $2$-spin model (see Sec. \ref{sec:spherical}) that $\b=\a-1$ and, consequently, the terms are of the same 
order. Therefore the {\it tilded} functions satisfy the following equation

\beq
\label{lead}
\begin{split}
&\tilde{\RR}(0) \int_0^1 dy\, y^{\a + \b} \tilde{\CC}(y) \\
+ &\tilde{\CC}(0) \int_0^1 dy\, (y^{\a} -1) y^{\b} \tilde{\RR}(y) \\
+ &\tilde{\CC}(0) \int_0^1 dy\,  [y^{\b} \tilde{\RR}(y) - \RR_{eq} (y)] \\
+ &\tilde{\CC}(0) \CC_{eq}(0) \\
+&\tilde{\RR}(0) \int_0^1 dy\, y^{\a -\b +\nu -2} \tilde{\CC}(y) =0
\end{split}
\eeq
The important point is that, if we separate the two terms, coming respectively from the order $\a - \nu$ and $\b - \nu +1$, 
and plug into Eq. (\ref{lead}) the exact solution for $\l = 0$ we find

\beq
\begin{split}
\EE_C^{(1)}\equiv &\tilde{\CC}(0) \int_0^1 dy\, (y^{\a} -1) y^{\b} \tilde{\RR}(y) \\
+ &\tilde{\CC}(0) \int_0^1 dy\,  [y^{\b} \tilde{\RR}(y) - \RR_{eq} (y)] \\
+ &\tilde{\CC}(0) \CC_{eq}(0)  \neq 0
\end{split}
\eeq
and
\beq
\begin{split}
\EE_C^{(2)}\equiv &\tilde{\RR}(0) \int_0^1 dy\, y^{\a + \b} \tilde{\CC}(y) \\
+&\tilde{\RR}(0) \int_0^1 dy\, y^{\a -\b +\nu -2} \tilde{\CC}(y) \neq 0
\end{split}
\eeq
If we reasonably assume that $\tilde{\CC}$ and $\tilde{\RR}$ change continuously form $\l=0$ to $\l\neq 0$, 
we know that in a certain neighborhood of $\l=0$ the true scaling functions would yield $\EE_C^{(1)} \neq 0$ and $\EE_C^{(2)} \neq 0$ 
so that the two equations cannot be satisfied separately and are necessarily of the same order. From this analysis 
we conclude that, independently of the value of $\l$, the two exponents satisfy 
\beq
\b = \a -1
\eeq
Given this first result, we can compute corrections to the leading behaviour.\\
\indent If we suppose that $\tilde{\CC}$ and $\tilde{\RR}$ admit a regular power-series 
expansion around $a=0$, namely 
\beq
\label{regular}
\begin{split}
&\tilde{\CC}(a)= \tilde{\CC}(0) \left[ 1+ \sum_{k=1}^{\infty} \tilde{\CC_k} a^k \right] \\
&\tilde{\RR}(a)= \tilde{\RR}(0) \left[ 1+ \sum_{k=1}^{\infty} \tilde{\RR_k} a^k \right] 
\end{split}
\eeq
than, with some effort, we can find that there are terms of order $2\a$ coming from the 
fifth and sixth line of Eq. (\ref{tildeeq}) that are equal in absolute value but opposite in sign, respectively 
\beq
\mp \, \tilde{\CC}(0) \tilde{\RR}(0) \int_0^1 \, dy \, y^{\a - 1}
\eeq
yielding a cancellation.
Therefore the equation at order $2\a$ would simply read
\beq
\label{absurd}
\frac{w_2}{w_1}\tilde{\CC}(0)^2=0
\eeq
that is satisfied only when $\l=w_2/w_1=0$, consistently with the fact that we know 
that Eq. (\ref{regular}) is true for $\l=0$. On the other hand, for any $\l \neq 0$, Eq. (\ref{absurd}) is 
intrinsically not satisfied, meaning that the hypothesis (\ref{regular}) is not verified in the general case. 
For this reason our Ansatz (\ref{trialC}) and (\ref{trialR}) is in principle incorrect, but still 
gives a quite accurate determination of the leading behaviour in $a \approx 0$ that is encoded in 
the exponent $\a$. \\
\indent For completeness we give Eq. (\ref{eq_scal_r}) written in terms 
of the {\it tilded} functions at leading order: 
\beq
\begin{split}
& \int_0^1 \left( y^{1-\a} -1 \right) y^{\nu +\a -2} \tilde{\RR}(y) \,dy \\ 
+&  \int_0^1 y^{\nu -1} \left[ y^{\a -1} \tilde{\RR}(y) -  \RR_{eq}(y)\right]  \,dy = 0
\end{split}
\eeq

\bibliography{paperbib.bib}
\bibliographystyle{unsrt}

\end{document}